%% file: NganguiaEtAl2020_arxiv.tex
\documentclass[aps, preprint,longbibliography,showpacs]{revtex4-1}
\usepackage{dcolumn}%
\usepackage{epsfig, graphics, amssymb, amsmath, color, bm, stmaryrd, tikz}
\usepackage{epstopdf}

\usepackage{lipsum}
\usepackage{color}
\usepackage{amsmath}
\usepackage{color}
\usepackage{siunitx}
\usepackage{float}
\usepackage{soul}
\usepackage{pifont,wasysym}
\usepackage{color,amsmath}
\usepackage{mathrsfs} 
\usepackage{amsmath}

\input{definitions}

\usepackage[unicode=true,
  linktocpage,
  linkbordercolor={0.5 0.5 1},
  citebordercolor={0.5 1 0.5},
  linkcolor=blue, citecolor = blue, colorlinks=true]{hyperref}

\definecolor{dgreen}{rgb}{0.0, 0.4, 0.0}

\usepackage{lipsum} 

\usetikzlibrary{shapes,backgrounds}

\newcommand{\sqdiamond}[1][fill=black]{\tikz [x=1.2ex,y=1.85ex,line width=.1ex,line join=round, yshift=-0.285ex] \draw  [#1]  (0,.5) -- (.5,1) -- (1,.5) -- (.5,0) -- (0,.5) -- cycle;}%
\newcommand{\MyDiamond}[1][fill=black]{\mathop{\raisebox{-0.275ex}{$\sqdiamond[#1]$}}}
\newcommand{\tikzcircle}[2][blue,fill=blue]{\tikz[baseline=-0.5ex]\draw[#1,radius=#2] (0,0) circle ;}
\newcommand\score[2]{%
  \pgfmathsetmacro\pgfxa{#1 + 1}%
  \tikzstyle{scorestars}=[star, star points=5, star point ratio=2.25, draw, inner sep=0.15em, anchor=outer point 3]%
  \begin{tikzpicture}[baseline]
    \foreach \i in {1, ..., #2} {
      \pgfmathparse{\i<=#1 ? "red" : "gray"} %
      \edef\starcolor{\pgfmathresult}
      \draw (\i*1em, 0) node[name=star\i, scorestars, fill=\starcolor]  {};
    }
    \pgfmathparse{#1>int(#1) ? int(#1+1) : 0}
    \let\partstar=\pgfmathresult
    \ifnum\partstar>0
      \pgfmathsetmacro\starpart{#1-(int(#1)}
      \path [clip] ($(star\partstar.outer point 3)!(star\partstar.outer point 2)!(star\partstar.outer point 4)$) rectangle 
      ($(star\partstar.outer point 2 |- star\partstar.outer point 1)!\starpart!(star\partstar.outer point 1 -| star\partstar.outer point 5)$);
      \fill (\partstar*1em, 0) node[scorestars, fill=yellow]  {};
    \fi
  \end{tikzpicture}%
}

\begin{document}

\title{Effects of Surfactant Solubility on the Hydrodynamics of a Viscous Drop in a DC Electric Field}

\author{Herve Nganguia$^1$, Wei-Fan Hu$^2$, Ming-Chih Lai$^3$ and Y.-N. Young$^{4}$}
\affiliation{$^1$Department of Mathematical and Computer Sciences,
Indiana University of Pennsylvania, Indiana, Pennsylvania 15705, USA\\
$^2$Department of Applied Mathematics, National Chung Hsing University, 145, Xingda Road, Taichung 402, Taiwan\\
$^3$Department of Applied Mathematics, National Chiao Tung University, 1001, Ta Hsueh Road, Hsinchu 30050, Taiwan\\
$^4$Department of Mathematical Sciences, New Jersey Institute of Technology, Newark, NJ 07102, USA}
\date{\today}

\thanks{Corresponding author: Y.-N. Young (yyoung@njit.edu)}  

\begin{abstract}
The physico-chemistry of surfactants (amphiphilic surface active agents) is often used to control the 
dynamics of viscous drops and bubbles. 
Surfactant sorption kinetics has been shown to play a critical role in the deformation of drops in extensional and 
shear flows, yet to the best of our knowledge these kinetics effects on a viscous drop in an electric field
have not been accounted for.
In this paper we numerically investigate the effects of sorption kinetics on a surfactant-covered viscous drop in an electric field. 
Over a range of electric conductivity and permittivity ratios between the interior and exterior fluids, 
we focus on the dependence of deformation and flow on the transfer parameter $J$, and Biot number $\text{Bi}$  
that characterize the extent of surfactant exchange between the drop surface and the bulk. 
Our findings suggest solubility affects the electrohydrodynamics of a viscous drop in distinct ways as 
we identify parameter regions where (1) surfactant solubility may alter both the drop deformation and circulation of fluid around a drop, 
and (2) surfactant solubility affects mainly the flow and not the deformation. 
\end{abstract}

 \maketitle
\section{Introduction}

Electric field is widely utilized to deform a viscous drop in microfluidics and many petroleum engineering applications. 
Electrohydrodynamics (EHD), generally referred to as the motion of fluid induced by an electric field, 
is highly relevant to transport and manipulation of small liquid drops in microfluidic devices.
Over the past two decades, dielectrophoresis, electro-osmosis,
and induced-charge electro-osmosis in EHD have deeply influenced the field of microfluidics. Moreover, the integration of EHD into microfluidic-based platforms has led to the development of technological platforms for 
manipulation of particles, colloids, droplets, and biological molecules across different length scales \citep{saville1970,ramos1998,ramos1999,castellanos2003,bazant2004,velev2006}.
EHD has been used in a wide range of applications, such as spray atomization, fluid motion of bubble
drop, electrostatic spinning, and printing \cite{hayati1986,ramos1994a,ramos1994b,saville1970,saville1993_PRL,Torza1971_PhilTransRSLa,sherwood1988}.
In material and bioengineering, EHD was utilized to manufacture nanostructured materials \cite{trau1996,trau1997}
and manipulate charged macromolecules \citep{vaidyanathan2015}.

For a leaky dielectric drop freely suspended in another leaky dielectric fluid, the bulk charge neutralizes on a fast timescale while ``free" charges accumulate on (and move along) the drop surface. 
In this physical regime, the full electrokientic transport model in a viscous solvent can be 
described by a charge-diffusion model that
can be further reduced to derive the Taylor-Melcher (TM) leaky dielecttric model \cite{Mori2018_JFM}.
In many physics and engineering applications with moderately dissolvable electrolytes,  
the TM leaky dielectric model can capture the deformation of a viscous drop in both dielectric medium \citep{OKonski1953_JChemPhys,Allan1962_ProcRSLa}
and a conducting medium \cite{Taylor1966_ProcRSLa2,Melcher1969_ARFM}.
The TM model has been extended in recent years to include the effects of charge relaxation \citep{Lanauze2013}, charge convection \citep{Lanauze2015,mandal2016_pre,mandal2017_pof,Das2017a}, and the investigation of non-spherical drop shapes \citep{Bentenitis2005_Langmuir,Zhang2012_PRE,Zabarankin2013_SIAM,Zabarankin2016_SIAM} and drop instabilities using direct numerical methods \citep{BrazierSmith1971_PhysFluids,BrazierSmith1971_ProcRSLa,Miksis1981_PhysFluids,Basaran1989_PhysFluids,Supeene2008_JCIS,Nganguia2014_CiCP,HuLaiYoung2014_JCP}.

In the absence of surface-active agent (surfactant), the balance between the electric stresses  and
 the hydrodynamic stress on the drop surface gives rise to a
 drop shape and a flow field that can be parametrized by
 the conductivity ratio and the permittivity ratio \citep{Saville1997_ARFM}.
Under a small electric field, a steady equilibrium drop shape exists due to the balance between the electric and hydrodynamic stresses \citep{Ha2000_JFM,Zholkovskij2002_JFM,Supeene2008_JCIS}. For a sufficiently large electric field, instabilities arise and the drop keeps deforming until it eventually breaks up into smaller drops \citep{Ha2000_PhysFluids,lh2007jfm}.
\begin{table}[htp]
\caption{Summary of published modeling work on the electrohydrodynamics (EHD) of a surfactant-laden viscous drop. SM denotes the small deformation (spherical harmonics) analysis, and LD refers to the large deformation (spheroidal harmonics) analysis. The abbreviations LS-RegM, BIM, and IIM stand for level-set regularized method, boundary integral method, and immersed interface method, respectively. Inertia-driven flow (Navier-Stokes) is shortened using N.-S.}
\vspace{-3mm}
\begin{center}
\begin{tabular}{p{2cm}p{3.cm}p{2cm}p{4.5cm}p{2.5cm}} \hline
Fluids & Electric field & Surfactants & Method & References \\ \hline\hline
Stokes & dc, uniform & insoluble & analytical (SD) & \citep{Ha1995_JCIS} \\
N.-S. & dc, uniform & insoluble & numerical (LS-RegM) & \citep{Teigen2010_PhysFluids} \\
Stokes & dc, uniform & insoluble & (semi-) analytical (LD) & \citep{Nganguia2013_PoF,Nganguia2018} \\
Stokes & dc, uniform & insoluble & numerical (BIM) & \citep{Lanauze2018sm,sorgentone18_jcp} \\
Stokes & dc, uniform & insoluble & analytical \& numerical & \citep{podar2018,podar2019a,podar2019b} \\
Stokes & dc, nonuniform & insoluble & analytical \& numerical & \citep{mandal2016_pof} \\
Stokes & dc, uniform & \textbf{soluble} & numerical (IIM) & Present Work \\
 \hline
\end{tabular}
\end{center}
\label{literature}
\end{table}%
Non-ionic surfactant has been extensively used for stability control
in experiments on electrodeformation of a viscous drop \citep{Ha1995_JCIS,OuriemiVlahovska14,zhang2015jpsc,Lanauze2018sm,luo2018cerd}.
By reducing  the surface tension and inducing a significant Marangoni stress due to the surfactant transport on the
interface, surfactant could lead to drastically different EHD of a surfactant-laden viscous drop. Table \ref{literature} summarizes the existing theoretical and numerical investigations in the literature.
In most of these studies \citep{Ha1995_JCIS,Teigen2010_PhysFluids,Nganguia2013_PoF,mandal2016_pof,Lanauze2018sm,sorgentone18_jcp,Nganguia2018,podar2018,sengupta2019_pre,podar2019a,podar2019b}, surfactants are assumed to be insoluble and the surface tension is described using either a linear relationship, or more realistically the Langmuir equation of state 
\begin{equation}\label{eq:langmuir}
\gamma(\Gamma) = \gamma_0 + RT\Gamma_\infty\ln\left(1 - \dfrac{\Gamma}{\Gamma_\infty} \right), 
\end{equation}
where $R$ and $T$ denote the gas constant and absolute temperature, respectively. 
$\gamma_0$ is the surface tension of an otherwise clean drop, 
and $\Gamma_\infty$ is the maximum surface packing limit.
A spheroidal model has been developed to predict the large electro-deformation of a viscous drop covered with
insoluble surfactant \cite{Nganguia2013_PoF}. Finite surfactant diffusivity has also been incorporated in such spheroidal
model \cite{Nganguia2018} with excellent agreement with full numerical simulations \cite{sorgentone18_jcp}.

Studies have shown that sorption kinetics and interactions between surfactants molecules can be effectively used to alter the concentration of surfactants at the drop interface \cite{chen1996,eggleton1998,BlawzdziewiczWajnrybLoewenberg1999_JFM,ZholkovskijKovalchuk2000_JCIS}, and have profound effects on the drop shape and dynamics \cite{MiLe94,HanyakSinzDarhuber2012_SoftMatt,LeRouxRocheCantatSaint-Jalmes2016_PRE,SellierPAnda2017_WaveMotion,ThieleArcherPismen2016_PRF,LiGupta2019_IndEngChemRes}. Electric field can in turn affect the rate of sorption kinetics \cite{sengupta2019_pre}.
These results naturally lead to the following inquiries: 
{\it What effects does adsorption and/or desorption have on EHD and how do they affect the interplay between all the various stresses?} To our knowledge  these questions have yet to be addressed in the literature.

In this work we aim to fill the gap by numerically solving the coupled equations for the leaky-dielectric model and surfactant transport equations. While our method is general enough to handle interaction between surfactants molecules, here we assume the relation provided by the Langmuir equation of state Eq. \ref{eq:langmuir} to focus on the effects of surfactants solubility.
In the present study, we investigate such dynamics in hopes of elucidating the physics governing the EHD of drops in the presence of soluble surfactants.

The paper is organized as follows: In \S \ref{s:formulation}, we present the physical problem and formulate the governing equations. 
Next, we present and discuss our findings: We consider the cases of low (\S \ref{s:results}) and high (\S \ref{s:fluxtransfer}) surfactants exchange between drop surface and bulk fluid. 
Finally, in \S \ref{s:conclusion} we end our study with a summary of how surfactants solubility affect the three modes of deformation (prolate `A', prolate `B', and oblate) for surfactant-covered drops in electric fields.

\section{Theoretical Modeling\label{s:formulation}}
\begin{figure}[t]
\begin{center}
\centerline{\includegraphics[keepaspectratio=true,width=2.5in]{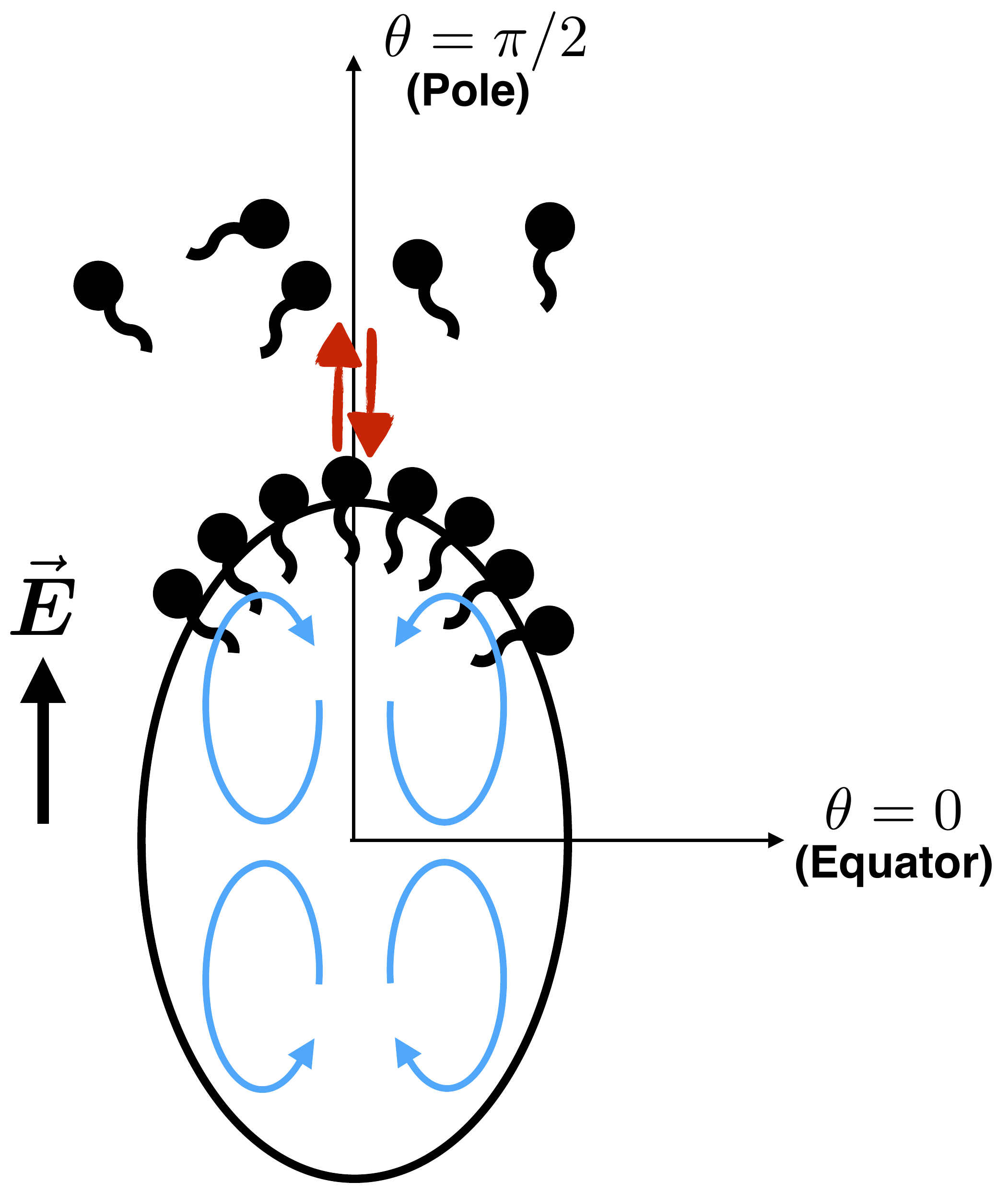}}
\caption{Sketch of the problem: A leaky dielectric viscous drop immersed in another dielectric fluid, with an external electric field $\vec{\textbf{E}}$ in the $z$ direction. The bead-rod particles represent insoluble and bulk surfactants. The double arrows denote adsorption-desorption kinetics while the curved arrows represent the induced flow.
}
\vspace{-7mm}
\label{fig1}
\end{center}
\end{figure}

We consider a viscous drop immersed in a leaky dielectric fluid in the presence of surfactants and subject to an electric field, as shown in figure \ref{fig1}. Each fluid is characterized by the fluid viscosity $\mu$, dielectric permittivity $\varepsilon$, and conductivity $\sigma$ with the superscript denoting interior (-) or exterior (+) fluid. In this work we denote the contrasts of those properties by $\mu_r = \mu^+/\mu^-$, $\varepsilon_r = \varepsilon^-/\varepsilon^+$, and $\sigma_r = \sigma^+/\sigma^-$.

\subsection{Formulation}
The fluids are governed by the incompressible Stokes equations, neglecting inertia
\begin{equation}
-\nabla p + \mu\nabla^2\boldsymbol{u} + \boldsymbol{F} = \boldsymbol{0},
\end{equation}
where $p$ and $\boldsymbol{u}$ are the pressure and velocity field, respectively; $\boldsymbol{F}$ is a singular force defined at the drop surface, as described below.
The electric field $\textbf{E} = -\nabla\phi$, where $\phi$ is the electric potential that satisfies the Laplace equation both inside and outside the drop in the extended leaky dielectric model,
\begin{equation}
\nabla^2\phi = 0.
\end{equation}
The surfactant transport on the drop surface and in the exterior bulk fluid are described by the following set of coupled equations
\begin{align}
\dfrac{\partial\Gamma}{\partial t} + \nabla_s\cdot\left(\Gamma\boldsymbol{v}_s\right) + \Gamma(\nabla_s\cdot\boldsymbol{n})\boldsymbol{u}_s\cdot\boldsymbol{n}  &= D_s\nabla_s^2\Gamma + \beta C_s\left(\Gamma_\infty - \Gamma\right) - \alpha\Gamma, \\
 \dfrac{\partial C}{\partial t} + \boldsymbol{u}\cdot\nabla C &= D\nabla^2C,
\end{align}
where $\boldsymbol{n}$ is the normal vector, $\boldsymbol{u}_s$ is the surface velocity on the drop and $\boldsymbol{v}_s = (I-\boldsymbol{n}\boldsymbol{n})\boldsymbol{u}_s$ is the velocity tangential component along the drop. $\Gamma$ and $C$ are the surfactant concentration on the drop surface and in the bulk, respectively; $C_s$ is the concentration of surfactant in the fluid immediately adjacent to the drop surface; $\alpha$ and $\beta$ are the kinetic constants for desorption and adsorption, respectively; $D_s$ and $D$ are the diffusion constant on the drop surface and in the bulk correspondingly.

At the drop interface, boundary conditions are imposed for the electric potential $\phi$, the flow field $\boldsymbol{u}$, and the bulk surfactant concentration $C$. First, the electric potential is continuous and the total current is conserved,
\begin{equation}\label{eq:qbc}
\llbracket \phi \rrbracket = 0, \quad \underbrace{\llbracket \sigma\nabla\phi\cdot\boldsymbol{n} \rrbracket }_{\text{Ohmic current}} = \underbrace{ \frac{d\tilde{q}}{dt}}_{\text{Charge relaxation}} + \underbrace{\nabla_s\cdot(\tilde{q} \boldsymbol{u}_s)}_{\text{Charge convection}},
\end{equation}
where $\tilde{q} = - \llbracket \varepsilon\nabla\phi\cdot\boldsymbol{n} \rrbracket$ represents the surface charge density, and $\llbracket \cdot \rrbracket$ denotes the jump between outside and inside quantities. 
The effects of charge relaxation on the transient behavior of drop \citep{Lanauze2013}, and of convection on equilibrium deformation \citep{Lanauze2015,Das2017a,Das2017b}, have been investigated analytically and numerically in the context of drops electrohydrodynamics. In the present study, we neglect these effects to more easily isolate the surfactant effects. This reduces Eq. \ref{eq:qbc} to  only consider the Ohmic current:
\begin{equation}
\llbracket \phi \rrbracket = 0, \quad \llbracket \sigma\nabla\phi\cdot\boldsymbol{n} \rrbracket  = 0.
\end{equation}
 Second, the electric and fluid problems are coupled through the stress balance
\begin{equation}\label{eq:stressbalance}
\underbrace{ \llbracket -p + \mu \left( \nabla\boldsymbol{u}^T + \nabla\boldsymbol{u} \right) \rrbracket }_{\text{Hydrodynamic stress}}\cdot\boldsymbol{n} + \underbrace{\llbracket \varepsilon \left( \boldsymbol{E}\boldsymbol{E} - \frac{1}{2}(\boldsymbol{E}\cdot\boldsymbol{E})\mathbf{I} \right) \rrbracket }_{\text{Electric stress}} \cdot\boldsymbol{n} = \underbrace{\gamma(\nabla_s\cdot\boldsymbol{n})\boldsymbol{n}}_{\text{Surface tension}} - \underbrace{\nabla_s\gamma }_{\text{Marangoni stress}}.
\end{equation}
Surfactants act to lower the surface tension, which now depends on the concentration of surfactants through the equation of state Eq. \ref{eq:langmuir}. As a result, the non-uniform surfactant distribution induced by the flow in and around the drop yields a surface tension gradient (the Marangoni stress). 

Finally, to close the system we need a third boundary condition that describes the flux of surfactants between the surface of the drop and and the bulk. The interfacial condition for the surfactant concentration,
\begin{equation}\label{eq:surfactantflux}
D \boldsymbol{n}\cdot\nabla C = \beta C_s\left(\Gamma_\infty - \Gamma\right) - \alpha\Gamma, 
\end{equation}
where $\boldsymbol{n}\cdot\nabla C = \partial C/\partial\boldsymbol{n}$ denotes the normal derivative of $C$.
We henceforth concentrate on axisymmetric solutions only.

\subsection{Nondimensionalization}
In this work we set the exterior fluid viscosity equal to the interior fluid viscosity: $\mu^+ = \mu^- = \mu$.
We use the drop size $r_0$ to scale length, capillary pressure $\gamma_0/r_0$ to scale pressure, equilibrium surfactant concentration $\Gamma_{eq}$ to scale $\Gamma$,
the far-field surfactant concentration $C_{\infty}$ to scale the bulk surfactant concentration, and electrically driven flow $U_d=\varepsilon^+ E_0^2r_0/\mu$ to scale velocity, in which $E_0$ denotes the intensity of the external electric field.

There are nine dimensionless physical parameters that characterize this system:
(1) the electric capillary number $\text{Ca}_E\equiv \mu U_d/\gamma_{eq} = \varepsilon^+ E_0^2r_0/\gamma_{eq}$
(ratio of electric pressure to capillary pressure), 
(2) permittivity ratio $\varepsilon_r = \varepsilon^-/\varepsilon^+$, 
(3) conductivity ratio $\sigma_r=\sigma^+/\sigma^-$,
(4) the elasticity constant $E= RT\Gamma_{\infty}/\gamma_0$ in the Langmuir equation of state, 
(5) the surfactant coverage $\chi=\Gamma_{eq}/\Gamma_\infty$, 
(6) the insoluble surfactant P\'eclet number $\text{Pe}_s=r_0U_d/D_s$, 
(7) the bulk surfactant P\'eclet number $\text{Pe}=r_0U_d/D$,
(8) the transfer parameter $J= C_\infty D/\Gamma_{eq} U_d$ and 
(9) the Biot number $\text{Bi}=\alpha\tau_{EHD}$ (ratio of EHD characteristic time scale to
desorption time scale).

The elasticity number $E$ measures the sensitivity of the surface tension to the surface surfactant concentration, whereas in the presence of surfactant exchange between the bulk and the drop interface, 
the surfactant coverage is related to the adsorption constant $k=\beta C_\infty/\alpha$ in Eq. \ref{eq:adsorbek}\cite{chen1996,eggleton1998}
\begin{equation}\label{eq:adsorbek} \chi =  \dfrac{k}{k+1}. \end{equation} 
The bulk and surface P\'eclet numbers denote the relative strength of convective transport versus diffusive transport. These two numbers also represent the ratio of two time scales: $\text{Pe} = \tau_D/\tau_{EHD}$, where $\tau_{EHD} = r_0/U_d$ is the EHD flow time scale, and $\tau_D = r_0^2/D$ is the surfactant diffusion time scale.
The parameter $J$ gives a measure of transfer of surfactant between its bulk and adsorbed forms relative to advection on the interface. It is important to note the ratio $\text{Bi}/J$ distinguishes two types of transport regime \citep{ChangFranses1995_CS,WangSiegelBooty2014_pof}: diffusion-controlled transport ($\text{Bi}/J>1$), and sorption-controlled transport ($\text{Bi}/J\ll1$).
In terms of the above dimensionless parameters, the clean drop cases correspond to $E=0$ or $\chi=0$ (Eq. \ref{eq:dimlessst}). The case of insoluble surfactants corresponds to $\text{Bi}=0$ (Eq. \ref{eq:dimensionlessflux}). The non-diffusive case corresponds to $\text{Pe}, \text{Pe}_s\gg1$.

We obtain the following dimensionless equations
\begin{align}
\label{eq:stokes}
-\nabla p + \text{Ca}\nabla^2\boldsymbol{u} + \boldsymbol{F} &= \boldsymbol{0}, \\
\nabla^2\phi &= 0,\\
\dfrac{\partial\Gamma}{\partial t} + \nabla_s\cdot\left(\Gamma\boldsymbol{v}_s\right) + (\nabla_s\cdot\boldsymbol{n})\boldsymbol{u}_s\cdot\boldsymbol{n} \Gamma  &= \dfrac{1}{\text{Pe}_s} \nabla_s^2\Gamma + J\boldsymbol{n}\cdot\nabla C, \label{eq:dimlessGamma} \\
  \dfrac{\partial C}{\partial t} + \boldsymbol{v}\cdot\nabla C  &= \dfrac{1}{ \text{Pe}}\nabla^2C, \\
 \gamma &= 1 + E \ln(1-\chi\Gamma)  \label{eq:dimlessst}
\end{align}
On the drop surface, the dimensionless boundary conditions are given by 
\begin{subequations}
\begin{align}
\label{eq:bc1}
& \llbracket \phi \rrbracket = 0,  \quad \llbracket \sigma\nabla\phi\cdot\boldsymbol{n} \rrbracket  = 0, \\
& \llbracket -p + \text{Ca} \left( \nabla\boldsymbol{u}^T + \nabla\boldsymbol{u} \right) \rrbracket \cdot\boldsymbol{n} + \llbracket \text{Ca}_E \left( \boldsymbol{E}\boldsymbol{E} - \frac{1}{2}(\boldsymbol{E}\cdot\boldsymbol{E})\mathbf{I} \right) \rrbracket \cdot\boldsymbol{n} = \gamma(\nabla_s\cdot\boldsymbol{n})\boldsymbol{n} - \nabla_s\gamma ,\label{eq:dimlessstressbalance} \\
& J\boldsymbol{n}\cdot\nabla C  = \text{Bi} \left[ C_s\left(1 + k -k\Gamma \right) - \Gamma \right].\label{eq:dimensionlessflux}
\end{align}
\end{subequations} 
In Eq.\ref{eq:stokes} and Eq. \ref{eq:dimlessstressbalance} the capillary number $\text{Ca}=\mu U_d/\gamma_0$ is the ratio of electric stress to tension in the absence of surfactant.
The singular force
\begin{equation}
\boldsymbol{F} = \int_0^{2\pi} \left( \boldsymbol{f}_\gamma + \text{Ca}_E\boldsymbol{f}_E \right)\delta^2\left( \boldsymbol{x} - \boldsymbol{X}(s,t)\right){\rm ~d}s,
\end{equation}
where
the electric force
\[
\boldsymbol{f}_E = \llbracket \varepsilon \left( \boldsymbol{E}\boldsymbol{E} - \frac{1}{2}(\boldsymbol{E}\cdot\boldsymbol{E})\mathbf{I} \right) \rrbracket \cdot\boldsymbol{n},
\]
and the surface tension force
\[
\boldsymbol{f}_\gamma =  \nabla_s\gamma - \gamma(\nabla_s\cdot\boldsymbol{n})\boldsymbol{n}.
\]
The right-hand side of the stress balance Eq. \ref{eq:dimlessstressbalance} shows that two surfactant-related mechanisms govern the deformation of drops. The first one is due to the non-uniform surfactant distribution that affects the surface tension. This mechanism acts in the normal direction, and is further broken down into two phenomena: tip-stretching and surface dilution \citep{pawar1996}. A measure of tip-stretching is the local surface tension for which $\gamma<1$ indicates {\it larger} deformation.  The area-average surface tension $\gamma_{\text{avg}}$ gives a global measure of the dilution effect; {\it smaller} deformations are attained for $\gamma_{\text{avg}}>1$. 
The second mechanism is driven by the Marangoni stress, which acts to suppress \citep{pawar1996,weidner2013_pof} or even reverse \citep{weidner2013_pof} surface convective fluxes. 
The Marangoni stress acts in the tangential directions, and consists of two principal components: the derivative of surface tension as a function of surfactant concentration ($\partial\gamma/\partial\Gamma$) and the surfactant concentration gradient ($\partial\Gamma/\partial \theta$), where $\theta$ is the angle parameter. 
These nontrivial and highly nonlinear mechanisms pose challenges in
studying the EHD of a surfactant-laden viscous drop.
Analytical solutions of the transport equation are only possible in very restricted limits \citep{Kallendorf2015pof}, 
and often numerical simulations are necessary. 
Several computational methods have been developed to simulate surfactants effects on 
droplets \citep{Tryg08,lai2011ijnam,xu2014cicp,xu2018jcp}. In the context of EHD, we refer the readers to the results in \citep{Teigen2010_PhysFluids,Lanauze2018sm,SorTor2018}.

In this work we implement a numerical code based on the immersed interface method (IIM) integrating numerical tools developed by our group \cite{Nganguia2014_CiCP,Nganguia2016_PRE,hu2018cf}. A description of the numerical setup is provided in \ref{appendA}, together with numerical validation in \ref{appendZ} and convergence study in \ref{appendB}. 

Moreover we fix the viscosity ratio $\mu_r=1$, the elasticity constant $E=0.2$ and conduct simulations with various combinations of parameters
to investigate the effects of surfactant solubility on the drop electrohydrodynamics. 
Our simulations show that deformation and flow patterns appear to be invariant with increasing surfactant solubility when the surfactant coverage $\chi<0.8$. We therefore focus our analysis on elevated surfactant coverage with $\chi=0.9$. \textcolor{black}{This surfactant coverage is in the relevant range in many experimental setups \cite{Ha1995_JCIS,Ha1998_JCIS,AM2006_PoF}, and the corresponding (dimensionless) surface tension $\gamma_{eq}=1+E\ln(1-\chi)=0.54$ and adsorption number $k=\chi/(1-\chi)=9$.}
The P\'eclet numbers $\text{Pe}=\text{Pe}_S=100$ for the oblate shapes (\S~\ref{ss:deformation})
and $\text{Pe}=\text{Pe}_S=500$ for the prolate shapes (\S~\ref{ss:transientA} and \S~\ref{ss:flow}).
These values of the P\'eclet numbers correspond to transfer parameter $J\ll1$; specifically $J=2\times10^{-3}$ for the prolate cases, and $J=10^{-2}$ for the oblate cases. This limit corresponds to the diffusion-controlled surfactant transport that is relevant in many practical applications \citep{WangSiegelBooty2014_pof}. Finally in \S~\ref{s:fluxtransfer} we investigate the
effect of solubility at larger values of $J$.

\section{Effects of surfactant physico-chemistry on drops electrohydrodynamics: $J \ll 1$\label{s:results}}

\begin{figure}[ht]
\centering
\includegraphics[keepaspectratio=true,width=5.in]{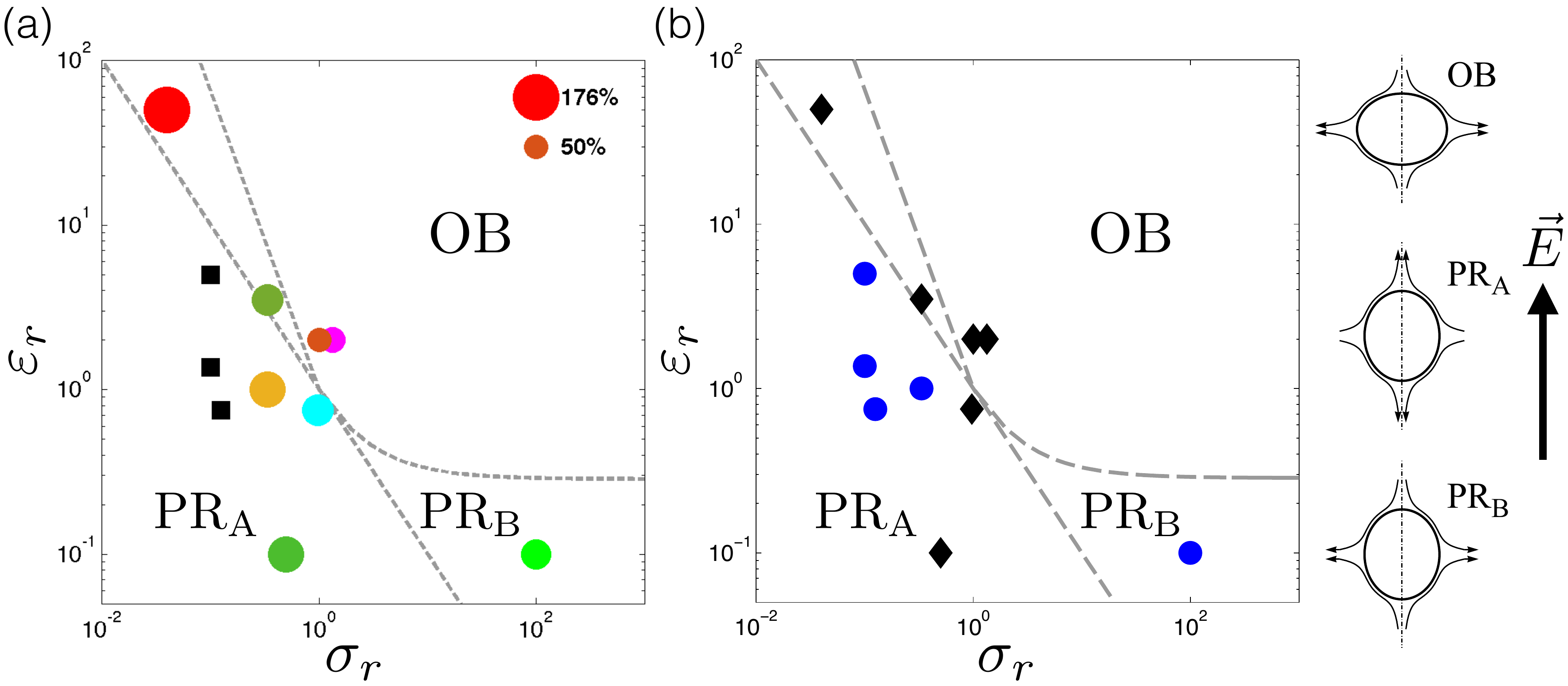}
%
\caption{\small $\sigma_r$--$\varepsilon_r$ phase plane depicting regions of prolate `A', prolate `B', and oblate drop shapes. The drops on the right-hand side depicts the expected circulation for each shape: counterclockwise (in the first quadrant) for the prolate `A', and clockwise for the prolate `B' and oblate shapes. 
(a) {\it Solubility effect on deformation}: ($\blacksquare$) indicates instability due to solubility, and the colored ($\bigcirc$) represent the relative change in deformation between clean and soluble drops. Larger sized circle point to greater solubility effect on deformation.
(b) {\it Solubility effect on flow}: The symbols denote the effects of surfactant solubility for a given $(\sigma_r,\varepsilon_r)$ pair on the flow in and around the drop:  
(\tikzcircle{2.5pt}) denotes a qualitative change in flow (reversal or stagnation point), and ($\MyDiamond$) represents no change in flow compared to the clean case. The electric capillary number $\text{Ca}_E = 0.25$, and the transfer parameter $J=10$.}\label{phaseplane}
\end{figure}
The shape of a viscous drop under an electric field
can be either prolate or oblate: A prolate shape is when a drop elongates along the applied electric field,
while an oblate shape is when a drop elongates in the orthogonal direction to the electric field.
Prolate drops are further categorized as `A' and `B', distinguished by the circulation patterns inside the drop: counterclockwise (equator-to-pole) for prolate `A', and clockwise (pole-to-equator) for prolate `B.' The oblate shape is characterized by a clockwise circulation (pole-to-equator). 

Figure~\ref{phaseplane} shows the phase diagram for the prolate versus oblate drops in the $(\sigma_r,\varepsilon_r)$ plane.  The dashed curves are boundaries for a clean (surfactant-free) viscous drop (see \cite{lh2007jfm} and references therein).
These boundaries are affected by insoluble surfactants \citep{Nganguia2013_PoF}, and this is corroborated by our 
current simulation results. 
Symbols in figure~\ref{phaseplane} are parameters collected from the literature.
\begin{table}[ht]
\caption{Surfactant effects at surfactant coverage $\chi=0.9$ and $J=10$. 
The circulation in the first quadrant is is either clockwise (C), counterclockwise (CC), or features a stagnation point (S) with eddies. }
\vspace{-5mm}
\begin{center}
\begin{tabular}{cccp{1.5cm}p{2.5cm}p{2.5cm}} \hline
$\sigma_r$ & $\varepsilon_r$ & Refs. & I (clean) &  II (insoluble)&  III (soluble)\\ \hline\hline
0.33 & 1 & \citep{Teigen2010_PhysFluids} & CC & C, stable & CC, stable  \\
0.33 & 3.5 & \citep{Teigen2010_PhysFluids} & C & C, stable & C, stable  \\
1 & 2 & \citep{Teigen2010_PhysFluids} & C & C, stable & C, stable  \\
0.1 & 5 & \citep{lh2007jfm} & CC & C, stable & CC, unstable  \\ 
0.04 & 50 & \citep{lh2007jfm} & C & C, unstable & C, stable  \\
100 & 0.1 & \citep{lh2007jfm} & C & S, unstable & C, stable  \\
0.1 & 1.37 & \citep{lh2007jfm,Ha2000_JFM} & CC & S, stable & CC, unstable  \\
0.97 & 0.75 & \citep{Lanauze2018sm} & CC & CC, stable & CC, stable \\ 
0.125 & 0.75 & \citep{Lanauze2018sm} & CC & S, stable & CC, unstable  \\
1.33 & 2 & \citep{mandal2016_pof} & C & C, stable & C, stable  \\
 \hline
\end{tabular}
\end{center}
\label{summary}
\end{table}%
We use these parameters to study the effects of surfactant solubility 
by drawing direct comparison with results for a clean drop. 

We summarize the solubility effects on drop deformation (with \text{Bi}=10) in figure~\ref{phaseplane}a, 
where the size of each circle correlates with the relative increase 
 in deformation between surfactant-covered and clean drops (with the smallest and the largest sizes in the legend).
Filled squares ($\blacksquare$) are for parameters where surfactant solubility gives rise to instability and the equilibrium shape (which exists for a clean drop) no longer exists.

\begin{figure}[ht]
\centering
\includegraphics[keepaspectratio=true,width=4.7in]{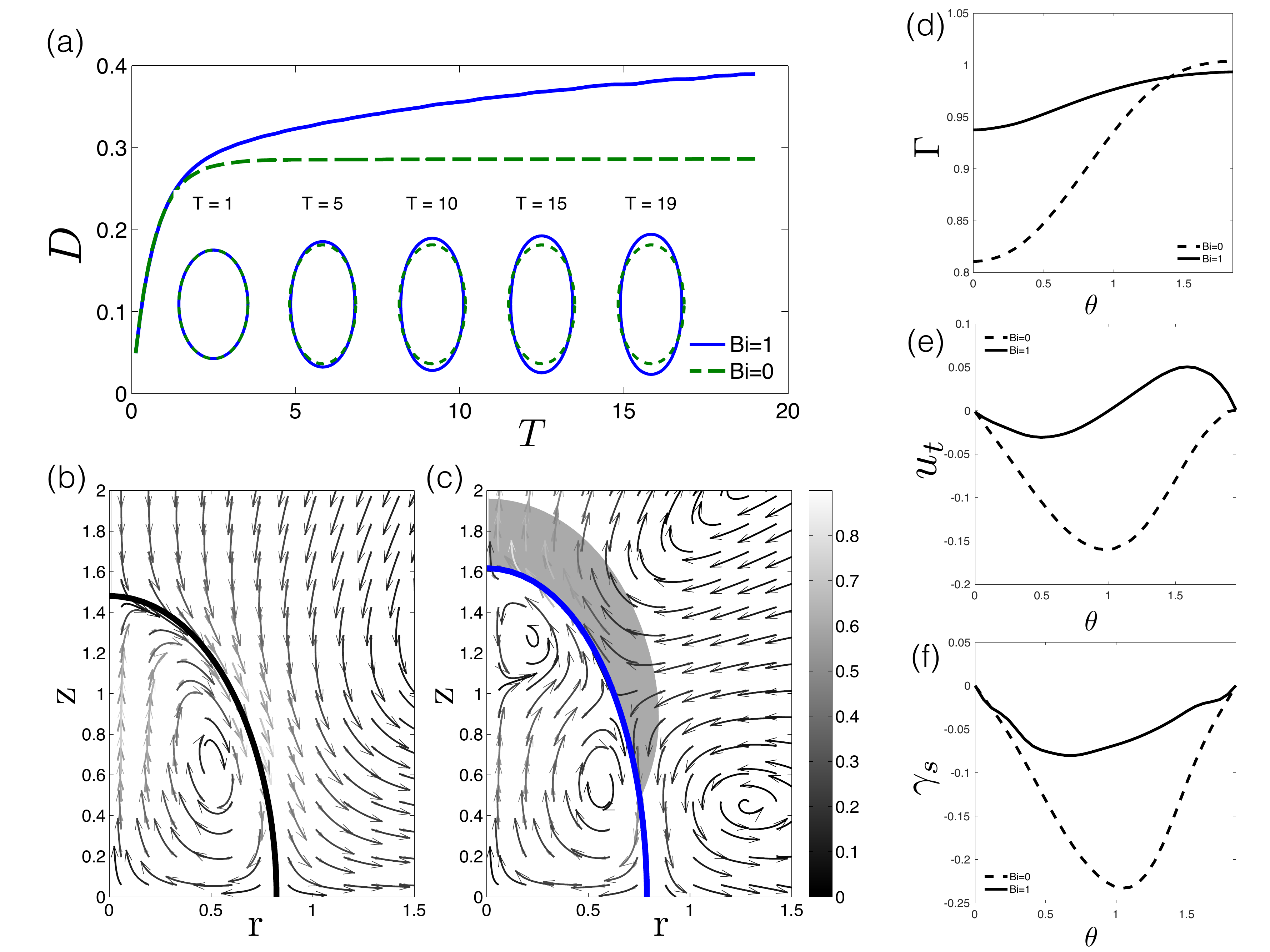}
%
\caption{\small 
A prolate drop with $(\sigma_r,\varepsilon_r)=(0.1, 5)$ and electric capillary number $\text{Ca}_E=0.25$.
(a) Deformation as a function of dimensionless time $T$ with Biot numbers $\text{Bi}=0$ (insoluble case) and $\text{Bi}=1$. The inset shows the drop shapes at various times. (b\&c) Flow field at $T=19$ for Biot number $\text{Bi}=0$ (b), and $\text{Bi}=1$ (c). (d) Surfactant distribution, (e) tangential velocity \textcolor{black}{$u_t = \boldsymbol{v}_s\cdot\boldsymbol{t}$}, and (f) Marangoni stress as a function of $\theta$, with the solid (dashed) curves for $\text{Bi}=0$ ($\text{Bi}=1$). In (c),  the 
surfactant sorption kinetics is found to be adsorption, color-coded by the blue on the drop surface.}\label{inserthere1}
\end{figure}

We summarize the solubility effects on the flow in figure \ref{phaseplane}b, 
where the filled diamonds ($\MyDiamond$) denote parameters where the flow around a surfactant-laden drop
is  qualitatively similar to the circulation pattern of a clean drop with the same parameters.
The blue circles \tikzcircle{2.5pt} represents parameters where the flow pattern is {\it qualitatively} changed by solubility.
For example, the flow inside a (\tikzcircle{2.5pt}) drop with $(\sigma_r,\varepsilon_r) = (0.1,5)$ changes from
a counterclockwise circulation (prolate `A') to a clockwise circulation (prolate `B') due to insoluble surfactant, then changes to a configuration of two counter rotating vortices due to surfactant solubility as shown in figure~\ref{inserthere1}.

Table~\ref{summary} summarizes the dynamics observed for each set of parameters. Column I is for the shape (circulation) of a clean drop: prolate `A' (counterclockwise, CC), prolate `B' (clockwise, C), and oblate (clockwise) (also see figure \ref{phaseplane}). 
Columns II and III are for the circulation inside a drop covered with insoluble surfactant ($\text{Bi}=0$) and soluble surfactant ($\text{Bi}=10$), respectively. 
Also in these two columns we indicate whether an equilibrium shape is attained (stable) or not (unstable) in the presence of insoluble (II) or soluble (III) surfactants at sufficiently high electric capillary number.
Below we elucidate the detailed solubility effects on the EHD of a viscous drop. Our simulations show that surfactant
effects are quite similar in regions of $(\sigma_r,\varepsilon_r)$ for prolate `B' and oblate  clean drop. Thus we focus
on regions where the clean drop is either prolate `A' or oblate.

\subsection{Increasing Biot number destabilizes a prolate drop \label{ss:transientA}}
Here we show that enhancing the surfactant solubility (by increasing Biot number) renders a prolate drop unstable.
Specifically we use the combination $(\sigma_r, \varepsilon_r) = (0.1, 5)$, where a surfactant-free viscous drop is
prolate `A' under an electric field. 
For a clean drop with $(\sigma_r, \varepsilon_r) = (0.1, 5)$ 
the steady equilibrium prolate `A' drop exists at all values of $\text{Ca}_E$ \citep{lh2007jfm}, 
whereas we establish equilibrium drop shape exists for the insoluble surfactant case up to $\text{Ca}_E=0.3$.
Figure \ref{inserthere1}a shows the transient deformation number as a function of dimensionless time $T$ for a prolate drop with $\text{Ca}_E=0.25$ and $\text{Bi}=1$.
Starting from a spherical drop covered with a uniform surfactant distribution both on the drop interface and in the bulk,
we simulate the drop EHD and examine the flow field, surfactant distribution, and 
the drop deformation number defined as
\begin{equation}\label{eq:Deq}
D = \frac{L-B}{L+B},
\end{equation}
where $L$ is the length of the major axis and $B$ is the length of the minor axis of the ellipsoid.

When the surfactant is insoluble ($\text{Bi}=0$) and weakly-diffusive (Peclet number $\text{Pe}_S\gg1$), 
the drop first elongates along the electric field with a flow from the equator to the pole, moving the surfactant from the equator to pole.
As the surfactant accumulates and builds up the Marangoni stress, the flow is reversed (from pole to equator) around $T\sim 0.6$ and
the drop reaches a equilibrium prolate shape with a clockwise circulation after
$T\sim 4$ in figure~\ref{inserthere1} in figure~\ref{inserthere1}b.
This circulation at equilibrium is opposite to that of a clean prolate `A' drop, and the flow magnitude is much smaller:
The Marangoni stress due to the non-diffusive insoluble surfactant changes the circulation from counter-clockwise (prolate `A' for the clean drop) to clockwise (prolate `B'). 

 However, as we increase the Biot number to allow for more surfactants exchange 
between the bulk and the surface of the drop, we find that the steady state no longer exists as 
the drop continuously deforms until the end of simulations (up to $T=20$) as illustrated in figure \ref{inserthere1}a.
 The surfactant distribution $\Gamma$, tangential velocity $\bm{u}_t$ and the Marangoni stress $\gamma_s$ at $T=20$
 are plotted in figures~\ref{inserthere1}d, e and f, respectively.
 
 For the case of insoluble surfactant ($\text{Bi}=0$) in figure~\ref{inserthere1}b,
  the Marangoni stress is able to sustain an equilibrium shape.
 In the simulations as we gradually increase the surfactant exchange between the bulk and the drop surface
 (by increasing $\text{Bi}$ from zero), 
 we find that the Marangoni stress is reduced in magnitude (figure~\ref{inserthere1}f) 
 because the surfactant on the drop surface is homogenized (figure~\ref{inserthere1}d)
 by the adsorption/desorption of surfactant. 

 Figure~\ref{inserthere1}e shows
 the corresponding tangential velocity on the drop interface.  We observe that the
 surfactant solubility not only reduces the magnitude of the tangential velocity but also gives rise to
 the development of counter rotating eddies inside the drop, as shown in figure~\ref{inserthere1}c.  
 Such counter rotating eddies inside a viscous drop
 are also observed in a clean viscous drop elongating indefinitely under a DC electric field \cite{lh2007jfm}.

\begin{figure}[htb]
\centering
\includegraphics[keepaspectratio=true,width=5.in]{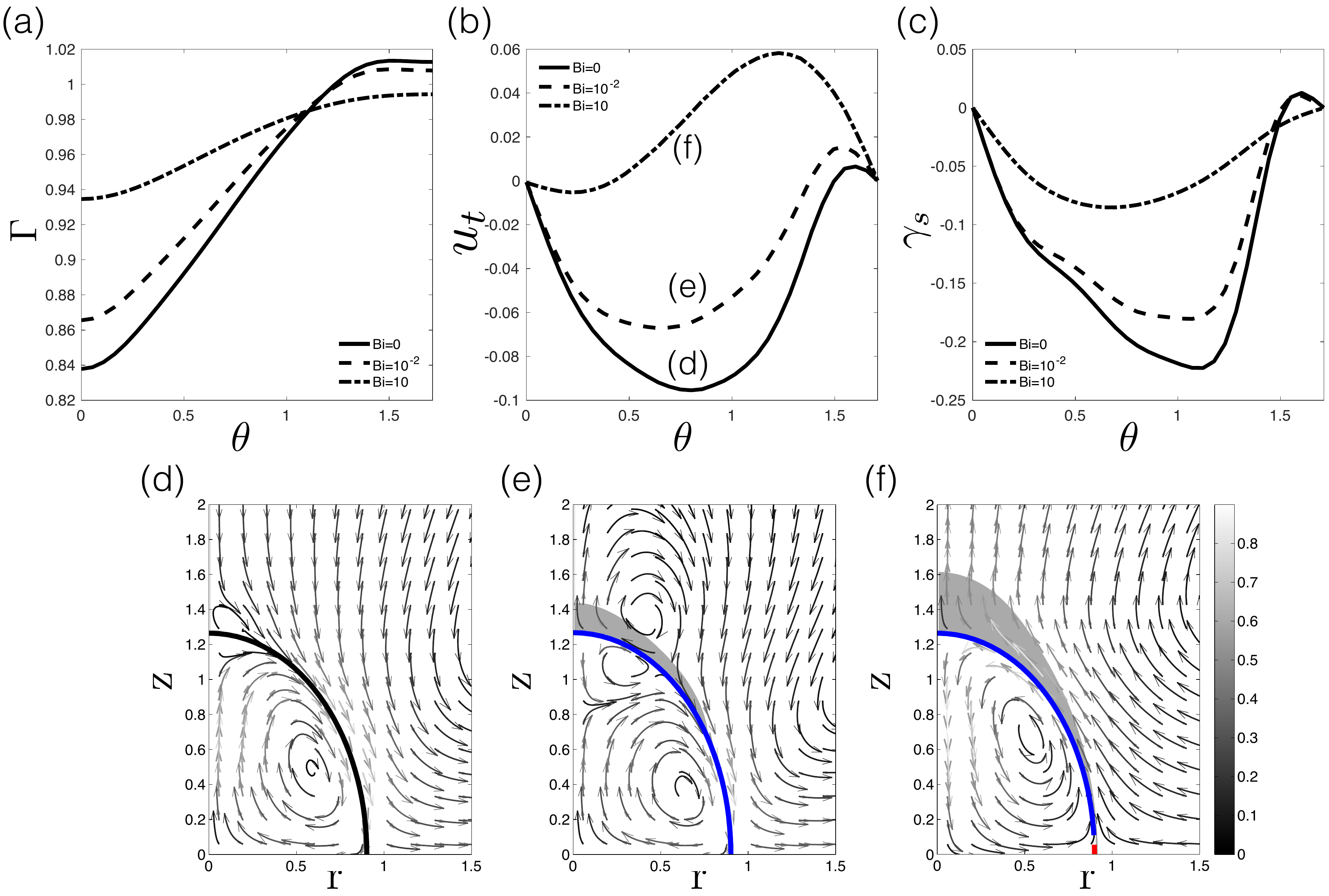}
%
\caption{\small 
A prolate drop with $(\sigma_r,\varepsilon_r) = (1/3,1)$ and $\text{Ca}_E=0.3$.
(a) Surfactant distribution, (b) tangential velocity $u_t = \boldsymbol{v}_s\cdot\boldsymbol{t}$, and (c) Marangoni stress as a function of $\theta$. 
$\text{Bi}=0$ for solid curves, $\text{Bi}=10^{-2}$ for dashed curves, 
and $\text{Bi}=10$ for dash-dotted curves. (d-f) Corresponding flow fields for (d) $\text{Bi}=0$, 
(e) $\text{Bi}=10^{-2}$ and (f) $\text{Bi}=10$. In (e)-(f),  the drop surface is color-coded to represent sorption kinetics: blue for adsorption, and red for desorption.}\label{fig22a}
\end{figure}

\subsection{Effects of Biot number on flow around a prolate drop\label{ss:flow}}
\begin{figure}[ht]
\centering
\includegraphics[keepaspectratio=true,width=4.5in]{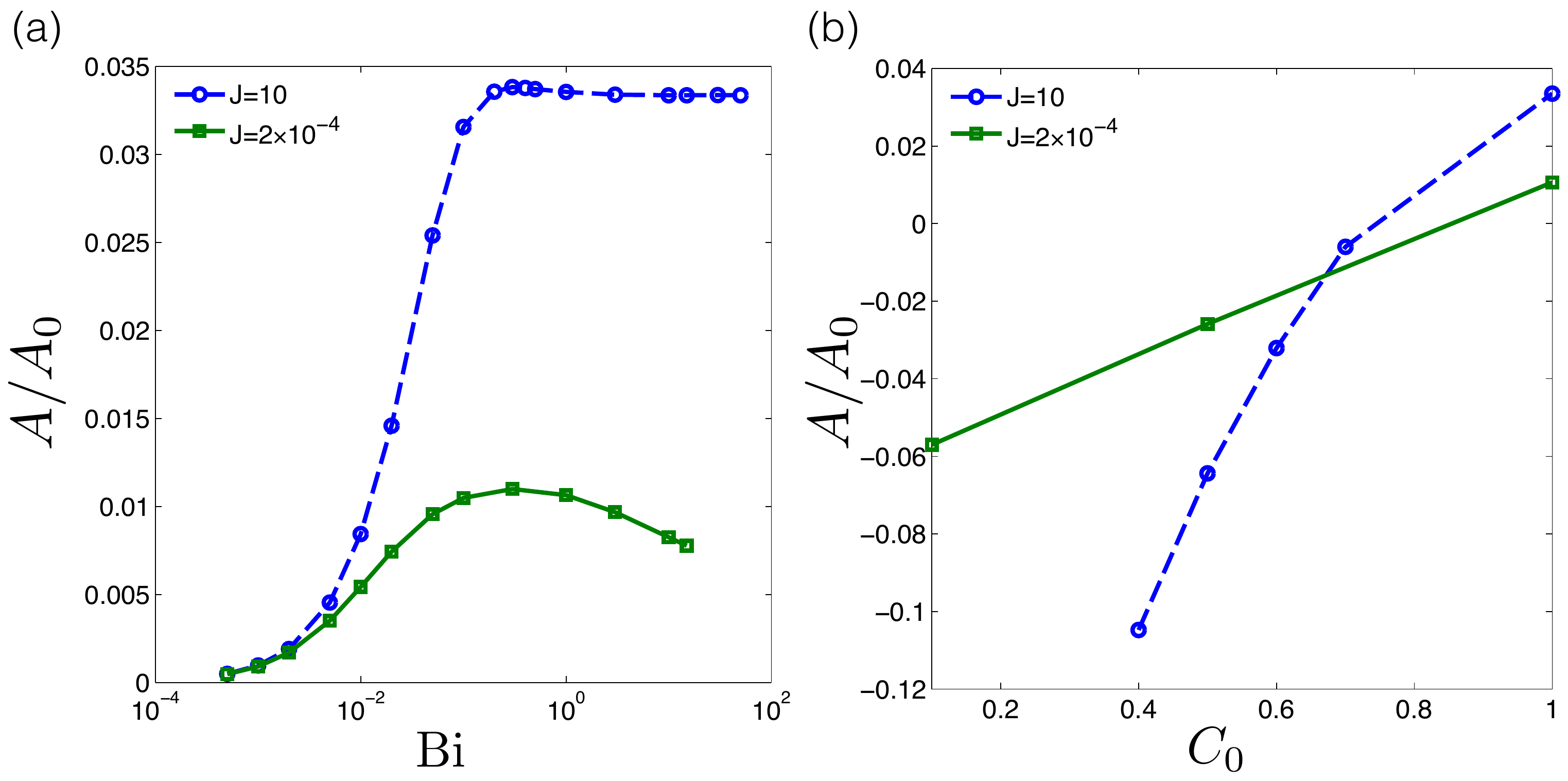}
%
\caption{\small 
A prolate drop with $(\sigma_r,\varepsilon_r) = (1/3,1)$ and $\text{Ca}_E=0.3$.
(a) Change in total amount $A$ of adsorbed surfactant for the prolate drop in \S \ref{ss:flow} with an initial bulk concentration $C_0=1$. 
(b) Change in total surfactant as a function of the initial bulk surfactant concentration, $C_0$. The Biot number $\text{Bi}=1$.}\label{fig24}
\end{figure}

In \S~\ref{ss:transientA} the surfactant solubility affects both drop deformation and the flow pattern of a prolate drop.
Here we investigate another scenario where the surfactant solubility affects only the flow pattern while the equilibrium 
drop shape remains close to the prolate shape 
of a drop covered with insoluble surfactants under an electric field.
Specifically we focus on the combination $(\sigma_r, \varepsilon_r) = (1/3, 1)$ with $\text{Ca}_E=0.3$.
Simulations show that the equilibrium drop deformation is minimally influenced
by surfactant solubility at all values of the electric capillary number
(with the change in deformation less than 1\% between the insoluble case and $\text{Bi}=10$) because sorption kinematics
induce little change in the total amount of surfactant as shown in figure~\ref{fig24}a (solid curve).  Consequently the
average surface tension does not vary much with $\text{Bi}$, leading to little change in drop deformation with increased surfactant solubility.

The flow pattern, on the other hand, is highly dependent on the surfactant distribution and kinetics.
Without surfactant the clean drop is prolate `A' at equilibrium with a counterclockwise flow under an electric field.
For $\text{Bi}=0$ the transport of an insoluble surfactant and the corresponding Marangoni stress gives rise to 
an interior flow dominated by a clockwise circulation with a small-counter rotating eddy around the pole as shown in
 figure \ref{fig22a}d, and the corresponding tangential velocity is shown in figure \ref{fig22a}b.
 As the Biot number is increased to $\text{Bi}=10^{-2}$ the counterclockwise eddy near the pole expands 
 as shown in figure \ref{fig22a}e, with the corresponding tangential velocity in figure \ref{fig22a}b.

When we further increase the Biot number ($\text{Bi}=10$), 
the counterclockwise eddy nearly takes over the whole interior flow (figure \ref{fig22a}f) as 
the surfactant is nearly constant (dash-dotted curve in figure~\ref{fig22a}a) and the Marangoni stress is of the smallest magnitude in figure~ \ref{fig22a}f.
This can be explained by examining the surface tension derivative and surfactant gradient. 
The former remains high, and strong Marangoni stresses are realized initially. 
However, adsorption dominates the surfactant exchange, and the surfactant distribution remains nearly uniform. 
This results in decreasing surfactant gradient, and therefore smaller overall Marangoni stress at equilibrium. 

To further examine how sorption/desorption of surfactant affects the drop deformation and flow pattern, 
we study the surfactant transport and distribution as follows.
First we investigate the total amount of adsorbed surfactant $A$, defined as the difference in total
amount of surfactant on the drop surface between the time $T$ and the initial time 0:
\begin{equation} 
A \equiv A_T - A_0 \equiv \int\Gamma(T){\rm~d}s - \int\Gamma(0){\rm~d}s. 
\end{equation}
Using this definition, $A>0$ denotes adsorption, and $A<0$ represents desorption.

 Figure \ref{fig24}a shows the total amount of adsorbed surfactant $A$ as a function of Biot number, for the prolate drop in \S \ref{ss:flow}. For an initial bulk surfactant concentration equals to the concentration in the far field, $A$ exhibits a non-monotonic behavior with a critical Biot number $\text{Bi}_{\text{cr}}\approx0.3$, where the adsorbed surfactant concentration is maximized. 
 \textcolor{black}{Moreover, we observe that adsorption ($A>0$) is the dominant kinetics for the full range of Biot number studied. Our simulations show this result is strongly dependent on the initial bulk surfactant concentration $C_0$. As illustrated in figure \ref{fig24}b that shows the total amount of adsorbed surfactant as a function of initial surfactant concentration in the bulk, desorption ($A<0$) becomes the dominant kinetics as $C_0$ is reduced.} 
Finally the stagnation point between two counter rotating eddies observed here are similar to those observed for multi-lobed, prolate-shaped clean drops \cite{lh2007jfm}. However, and unlike the case of clean drops in \cite{lh2007jfm}, we hypothesize the flow reversal and eddies formation are driven by competition between the electrically-induced and Marangoni flows, possibly in similar manner as reported in previous findings on surfactant-laden liquid films under gravity \cite{weidner2013_pof}.
\begin{figure}[ht]
\centering
\includegraphics[keepaspectratio=true,width=4.7in]{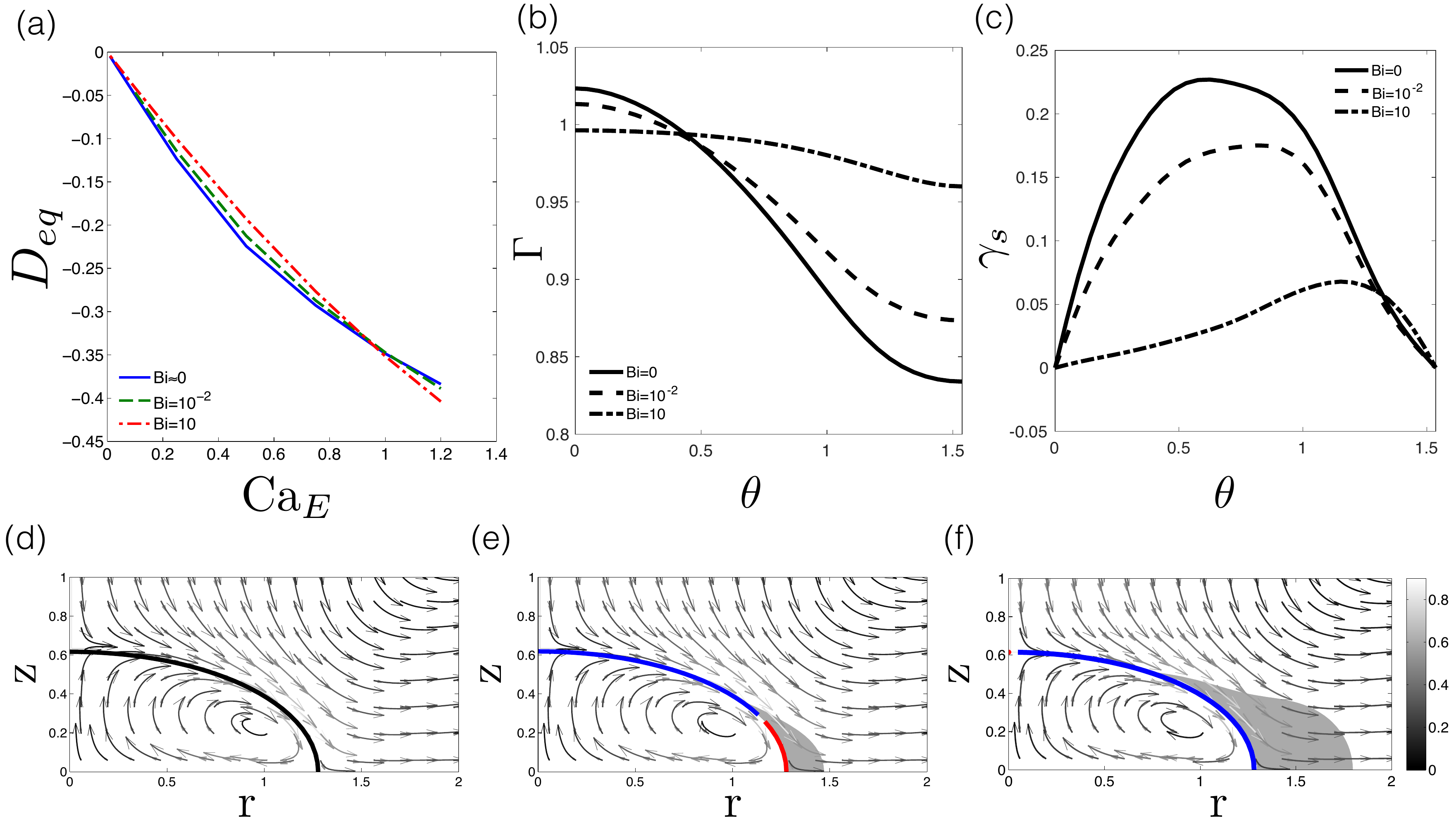}
%
\caption{\small An oblate drop with $(\sigma_r,\varepsilon_r) = (1,2)$. 
(a) Deformation number as a function of $\text{Ca}_E$. 
(b) Surfactant distribution, and (c) Marangoni stress as a function of $\theta$. Solid lines are for $\text{Bi}=0$ (the insoluble surfactant case), 
dashed lines are for $\text{Bi}=10^{-2}$, and dash-dotted lines are for $\text{Bi}=10$.
(d-f) Flow field at $\text{Ca}_E=1$ for Biot number $\text{Bi}=0$ (d), $\text{Bi}=10^{-2}$ (e), and $\text{Bi}=10$ (f).
In (d-f)  the drop surface is color-coded to represent sorption kinetics: blue for adsorption, and red for desorption.}\label{fig8b}
\end{figure}
\subsection{Effects of Biot number on equilibrium deformation of an oblate drop\label{ss:deformation}}
Here we consider the combination $(\sigma_r, \varepsilon_r) = (1, 2)$ that corresponds to a surfactant-laden oblate drop (with and without surfactant solubility).
The equilibrium deformation shows a visible dependence on both the electric capillary and Biot numbers, 
as illustrated in figure \ref{fig8b}a. The deformation undergoes a transition around $\text{Ca}_E\approx1$: 
The absolute deformation is smaller at low to moderate electric field strength compared to the insoluble case, 
while increasing the Biot number yields larger deformation at electric capillary numbers $\text{Ca}_E>1$.

Figures \ref{fig8b}b\&c show the surfactant distribution and Marangoni force as a function of $\theta$ at $\text{Ca}_E=1$. 
The corresponding flow field and the bulk surfactant distribution are 
in figures \ref{fig8b}d ($\text{Bi}=0$), \ref{fig8b}e ($\text{Bi}=10^{-2}$), and \ref{fig8b}f ($\text{Bi}=10$). 
For $(\sigma_r, \varepsilon_r) = (1, 2)$ we find that the interior flow remains a clockwise circulation (from pole to equator)
for all values of the Biot number.

Locally at the equator ($\theta=0$), the surface tension is less than $\gamma_{eq}$ for insoluble and weak surfactant exchange, 
suggesting that tip-stretching dominates. However, at higher Biot number the surface tension at the equator is slightly greater than $\gamma_{eq}$. 
Looking at sorption kinetics, adsorption dominates but for a region of desorption near the equator ($\text{Bi}=10^{-2}$ in figure~\ref{fig8b}e), 
while adsorption dominates on the entire drop surface for strong surfactant exchange ($\text{Bi}=10$ in figure~\ref{fig8b}f).  
In terms of surface dilution, the average surface tension $\gamma_{\text{avg}}$ remains less than unity 
with increasing Biot number. This couples with the local surface tension at the equator that is 
above $\gamma_{eq}$, suppressing deformation. 

\begin{figure}[ht]
\centering
\includegraphics[keepaspectratio=true,width=4.7in]{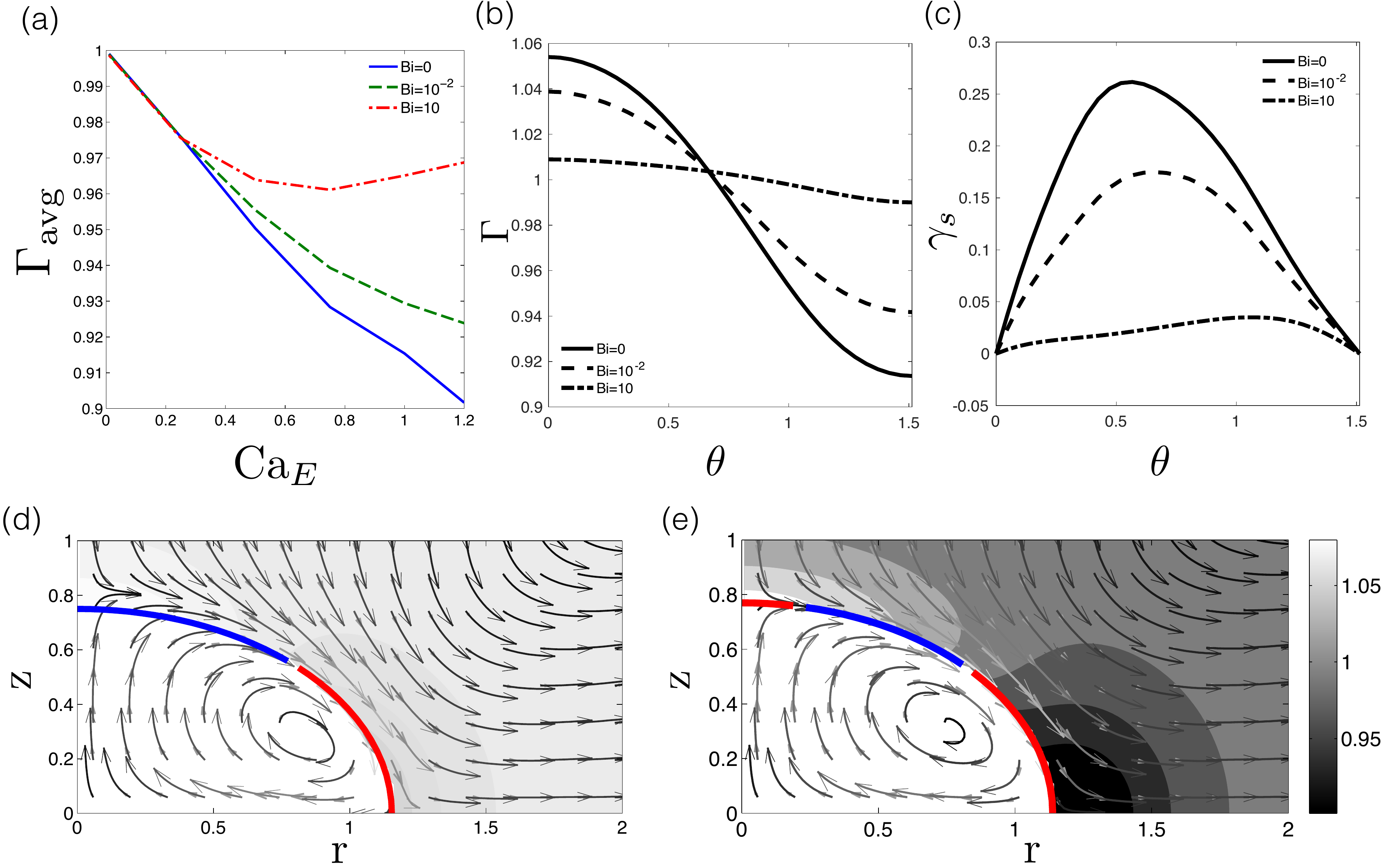}
%
\caption{\small An oblate drop with $(\sigma_r,\varepsilon_r) = (1,2)$. 
(a) Average surfactant concentration as a function of $\text{Ca}_E$. 
(b) Surfactant distribution, and (c) Marangoni stress as a function of $\theta$ at $\text{Ca}_E=0.5$. 
Solid lines are for $\text{Bi}=0$ (the insoluble surfactant case), 
dashed lines are for $\text{Bi}=10^{-2}$, and dash-dotted lines are for $\text{Bi}=10$.
(d-e) Flow field at $\text{Ca}_E=0.5$ for Biot number $\text{Bi}=10^{-2}$ (d) and $\text{Bi}=10$ (e).
In (d-e)  the drop surface is color-coded to represent sorption kinetics: blue for adsorption, and red for desorption.}\label{fig6_addentum}
\end{figure}
Figure~\ref{fig6_addentum}a shows the average surfactant $\Gamma_{\text{avg}}$ as a function of $\text{Ca}_E$. 
The rise of $\Gamma_{\text{avg}}$ for $\text{Ca}_E > 0.8$ for
$\text{Bi}=10$  corresponds to a reduced capillary pressure associated with the enhanced drop deformation in figure~\ref{fig8b}a.
Figure~\ref{fig6_addentum}b\&c show the surfactant distribution and Marangoni stress at \textcolor{black}{$\text{Ca}_E=0.5$},
and the corresponding flow field in figure~\ref{fig6_addentum}d ($\text{Bi}=10^{-2}$) and figure~\ref{fig6_addentum}e ($\text{Bi}=10$).

\section{Effects of surfactant physico-chemistry on drops electrohydrodynamics: $J >1$\label{s:fluxtransfer}}
As we specified earlier, the ratio $\text{Bi}/J$ differentiates between diffusion-controlled transport ($\text{Bi}/J>1$), and sorption-controlled transport ($\text{Bi}/J\ll1$). 
In the previous section we focus on the diffusion-controlled regime.
Here we focus on the sorption-controlled regime (with $J=10$) and make comparison with results for the diffusion-controlled regime in \S\ref{s:results}.

\subsection{Unstable drop dynamics\label{subsec:unstable_drop_dynamics_largeJ}}
\begin{figure}[h!]
\centering
\includegraphics[keepaspectratio=true,width=5.in]{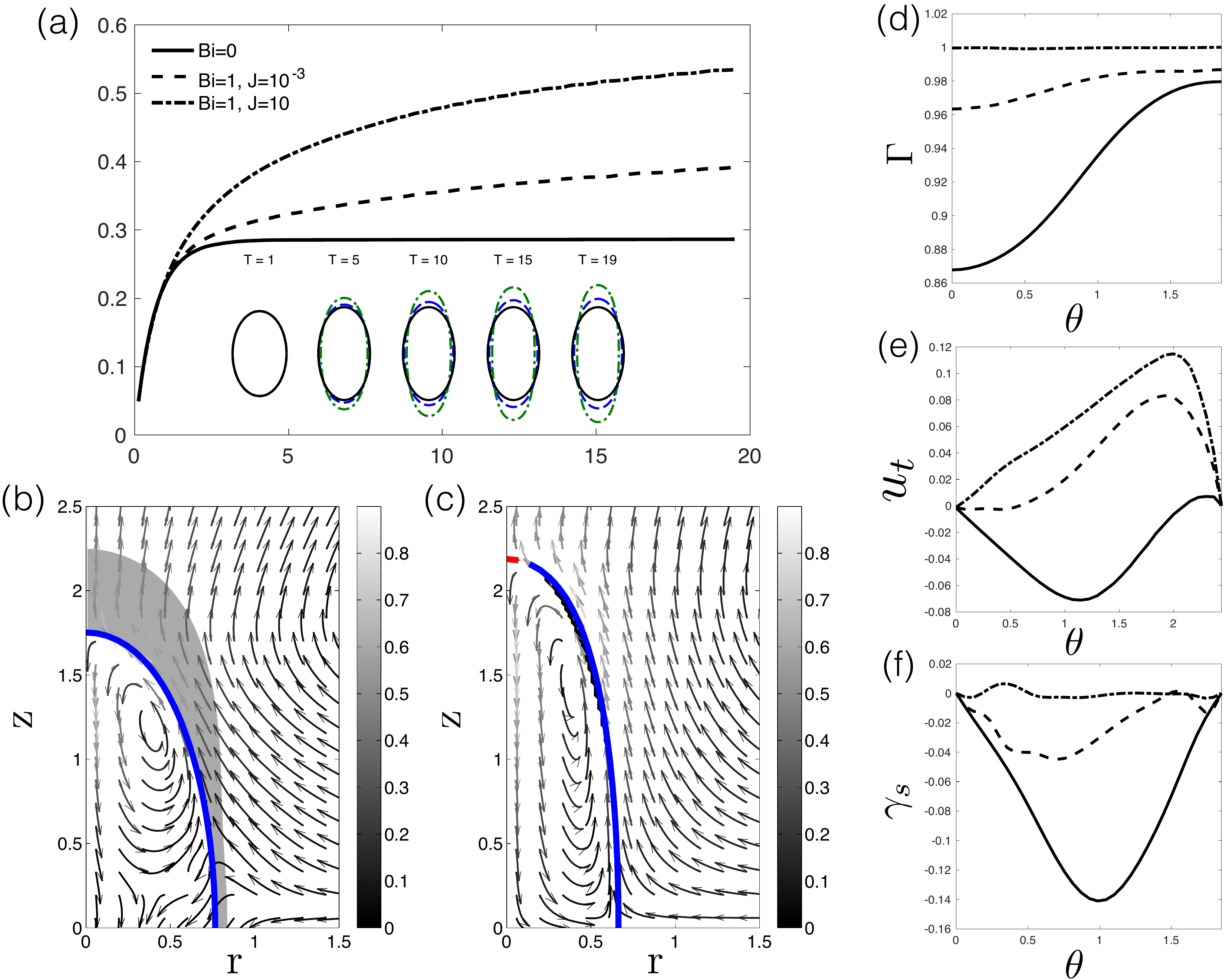}
%
\caption{\small 
A prolate drop with $(\sigma_r,\varepsilon)=(0.1,5)$ and $\text{Ca}_E=0.25$.
(a): Drop shapes for the prolate drop. (b) and (c): Flow field at $T=19$ for $J=10^{-3}$ (b) and $J=10$ (c) with $\text{Bi}=1$.
(d): Surfactant distribution, (e): tangential velocity \textcolor{black}{$u_t = \boldsymbol{v}_s\cdot\boldsymbol{t}$}, and (f): Marangoni stress. The solid lines are for $\text{Bi}/J=0$. 
The dashed and dash-dotted lines are for $\text{Bi}=1$ with $J=10^{-3}$ and $J=10$, respectively. 
In (b-c)  the drop surface is color-coded to represent sorption kinetics: blue for adsorption, and red for desorption.}\label{inserthere7}
\end{figure}
First we focus on the prolate drop with $(\sigma_r,\varepsilon_r)=(0.1, 5)$ (\S \ref{ss:transientA})
and make comparison between $J=10^{-3}$ (figure~\ref{inserthere1}) and $J=10$ with $\text{Bi}=1$.
Figure \ref{inserthere7}a shows the drop shape from $T=1$ to $T=19$. Figure \ref{inserthere7}b\&c are  the corresponding flow field at $T=19$ with $J=10^{-3}$
and $J=10$, respectively.
For insoluble surfactants ($\text{Bi}/J=0$, solid curves in figure~\ref{inserthere7}a, d, e \& f), 
the surfactant has the most spatial inhomogeneity that corresponds to a large Marangoni
stress. With soluble surfactant in the diffusion-controlled regime ($\text{Bi}/J > 1$, dashed curves) 
the surfactant sorption kinetics greatly reduces the Marangoni stress, giving rise to larger drop deformation.
In the sorption-controlled regime ($\text{Bi}/J = 0.1 <1$, dash-dotted curves) the surfactant concentration $\Gamma$ is
nearly homogeneous and the Marangoni stress is quite small, corresponding to the largest and fastest deformation in (a).
We note that suppressing the Marangoni stress in the diffusion-controlled regime gives rise to a $25\%$ increase in drop deformation (compared to the insoluble case), while in the
sorption-controlled regime a $60\%$ increase in drop deformation is found in the simulations.

Finally we observe that for the sorption-controlled case, the surfactant kinetics at the drop tip ($\theta=\pi/2$ in figure~\ref{inserthere7}c) is dominated by desorption
 (red portion of the drop surface in figure~\ref{inserthere7}c) while for the diffusion-controlled case the surfactant 
 kinetics is dominated by adsorption all over the drop (see blue portion of the drop surface in  figure~\ref{inserthere1}c). However, the total amount of surfactant
 increases on the drop surface for both cases.

\subsection{Transient overshoot and equilibrium drop dynamics\label{subsec:transient_overshoot}}
\begin{figure}[h!]
\centering
\includegraphics[keepaspectratio=true,width=5.in]{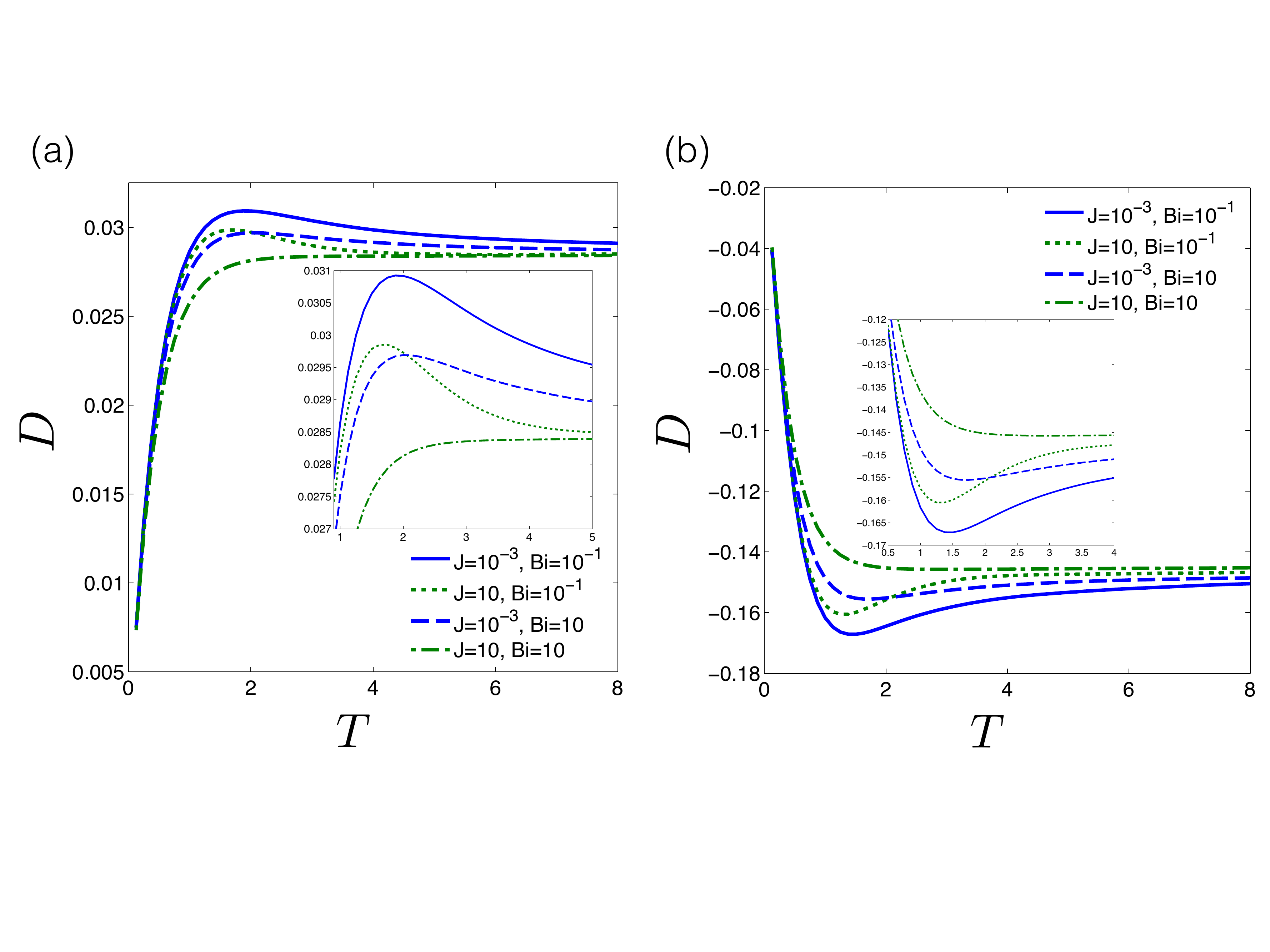}
%
\caption{\small 
Deformation as a function of dimensionless time $T$ for (a) a prolate  drop with $(\sigma_r,\varepsilon)=(0.97,0.75)$, 
and (b) the oblate drop with $(\sigma_r,\varepsilon_r)=(4/3,2)$. 
$\text{Ca}_E=0.25$ for both cases. The solid and dotted lines are for low ($J=10^{-3}$), and high ($J = 10$) transfer parameter with $\text{Bi}=10^{-1}$, respectively. The dashed and dash-dotted lines represent higher Biot number ($\text{Bi}=10$) with $J=10^{-3}$ and $J=10$, respectively. }\label{inserthere8}
\end{figure}
In our simulations we observe that 
the transient dynamics of drop deformation depends on $J$: Figure~\ref{inserthere8} shows that, at a given value of $\text{Bi}$, the drop deformation
number $D$ displays an overshoot en route to the equilibrium for small $J$. 
Such overshoot in the drop deformation is found for weakly diffusive insoluble surfactant \citep{Nganguia2018}.
However, as shown in figure \ref{inserthere8} (see inset for close-up of the transient overshoot),
the transient overshoot dynamics is suppressed at large $J$: In this case, the deformation monotonically reaches its equilibrium value. 
We note these observations are valid for both prolate (figure \ref{inserthere8}a) or oblate (figure \ref{inserthere8}b) drops.

\begin{figure}[ht]
\centering
\includegraphics[keepaspectratio=true,width=4.5in]{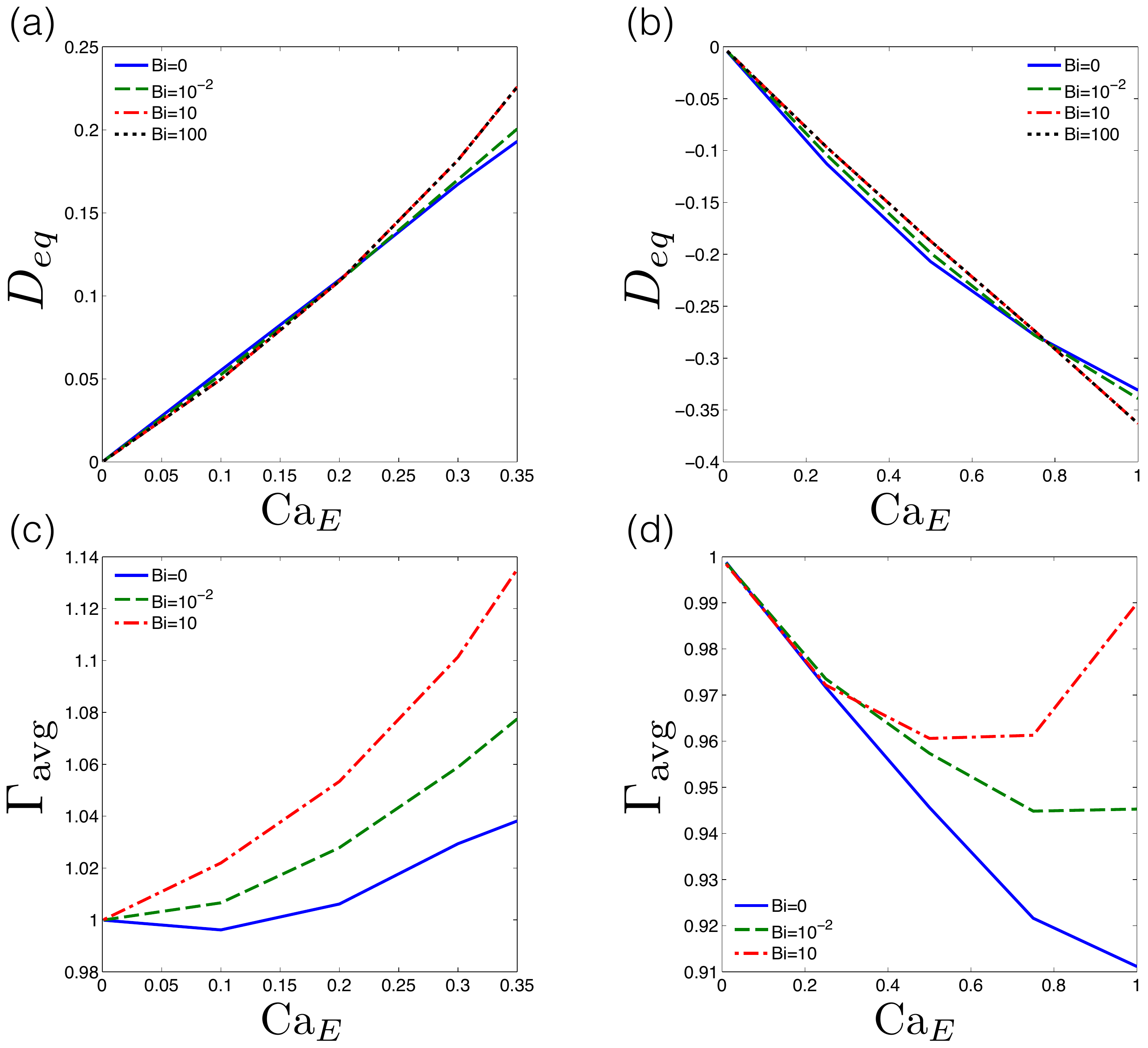}
%
\caption{\small 
Equilibrium drop deformation and average surfactant coverage as a function of electric capillary number $\text{Ca}_E$ for four values of $\text{Bi}$ (as labeled) and $J=10$. 
(a) and (c): a prolate drop with $(\sigma_r,\varepsilon)=(1/3,1)$.
(b) and (d): an oblate drop with $(\sigma_r,\varepsilon_r)=(1,2)$.
 }\label{inserthere5}
\end{figure}

\begin{figure}[ht]
\centering
\includegraphics[keepaspectratio=true,width=5.in]{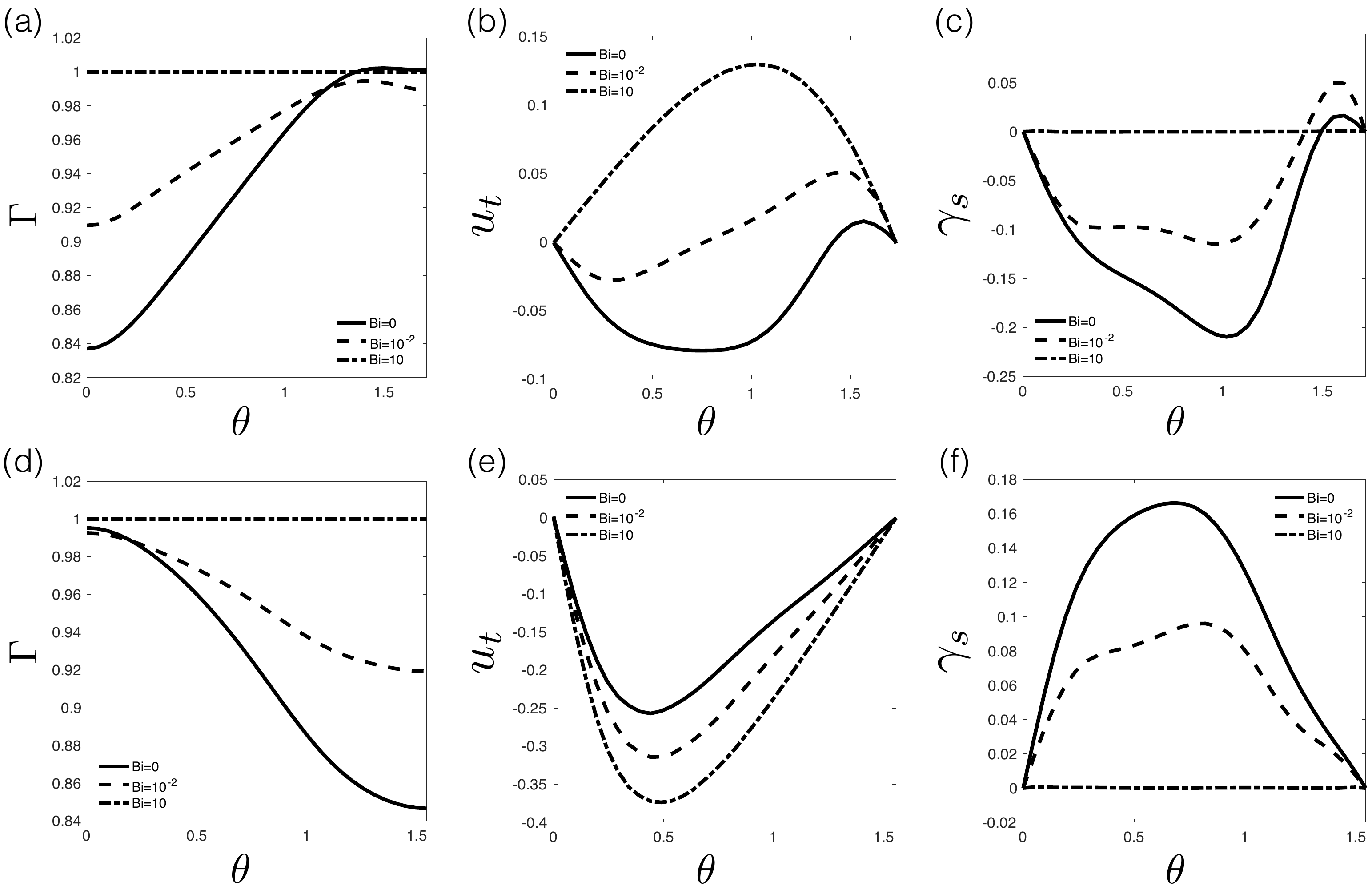}
%
\caption{\small 
Surfactant distribution (first column), tangential velocity (\textcolor{black}{$u_t = \boldsymbol{v}_s\cdot\boldsymbol{t}$}, second column), and Marangoni force (third column) for 
a prolate drop (a-c) with $\text{Ca}_E=0.35$, and an oblate drop (d-f) with $\text{Ca}_E=1$
 in figure \ref{inserthere5}. The corresponding flow fields for each shape and varying Biot numbers are shown in \ref{appendJ}.
 }\label{inserthere6}
\end{figure}

Figure \ref{inserthere5} shows the equilibrium deformation as a function of electric capillary number 
for a prolate drop with $(\sigma_r,\varepsilon_r)=(1/3,1)$ (figure \ref{inserthere5}a) and an oblate drop with $(\sigma_r,\varepsilon_r) = (1,2)$
  (figure \ref{inserthere5}b) at $J=10$.
Figure~\ref{inserthere5}c\&d show the corresponding average surfactant concentration $\Gamma_{\text{avg}}$ versus $\text{Ca}_E$.
With $J=10$, we expect the Biot number to play a more significant role in the deformation of the drop. 
This is especially true for the prolate drop, and is reflected in figure \ref{inserthere5}a. 
While  the drop deformation for a prolate drop in figure~\ref{inserthere5}a 
does not depend much on $\text{Bi}$ for $\text{Ca}_E<0.2$, 
solubility effects become significant for $\text{Ca}_E > 0.2$. 
At $\text{Ca}_E=0.4$, the equilibrium drop deformation for $\text{Bi}=10$ is more than 20\% larger  than that of the insoluble case ($\text{Bi}=0$).

For $\text{Bi} \ge 10$ we find that 
the equilibrium drop deformation does not depend on surfactant solubility again.
This is because the surfactant transport transitions from sorption-controlled to  diffusion-controlled dynamics as 
we increase from $\text{Bi}=10^{-2}$ to $\text{Bi}=10$ with $J=10$.
In the diffusion-controlled regime, the drop deformation dynamics is discussed in \S\ref{s:results} (where $J\ll 1$).
Once in the sorption-controlled regime $\text{Bi}/J \ll 1$ the surfactant on the drop surface is highly homogenized and thus the deformation is dominated by the 
balance between the normal Maxwell stress and the normal hydrodynamic stress.

In figure~\ref{inserthere6} we show how $\text{Bi}$ affects the spatial variation of 
the surfactant distribution (a\&d), tangential velocity (b\&e) and Marangoni stress (c\&f)
for the two sets of $(\sigma_r, \varepsilon_r)$ with $\text{Ca}_E=0.35$ for the prolate case and $\text{Ca}_E=1$ for the oblate case.
Overall we find qualitative similarity in the effects of $\text{Bi}$ between $J=10$ and $J\ll1$ in \S~\ref{s:results}:
For a prolate drop (figures \ref{inserthere6}d, \ref{insertappend1}), increasing the Biot number transitions the flow from a complete reversal for $\text{Bi}=0$
to development of counter rotating eddits, and then back to its {\it natural} prolate `A' circulation for a surfactant-free drop. 
On the other end, an oblate drop (figures \ref{inserthere6}e, \ref{insertappend3}) maintains the same clockwise circulation with increasing $\text{Bi}$.

At high value of the Biot number ($\text{Bi}\geq10$), figures \ref{inserthere6}a\&d show surfactants are uniformly distributed over the drop surface. 
In this case, high values of the Biot number and transfer parameter combine to produce uniform surfactant distributions and the drop behaves as if it is a
surfactant-free drop with a much reduced surface tension. 
This is similar to the diffusion-dominated regime ($\text{Pe}_S=0$) of a viscous drop covered with insoluble surfactant ($\text{Bi}=0$)
\citep{Nganguia2018}.

We also found that at $J=10$ the sorption kinetics depends on the drop shape: Adsorption of surfactant occurs on the surface of a prolate drop  (figure~\ref{insertappend1})
while for an oblate drop desorption takes place around the equator (figure~\ref{insertappend3}).
This in turn increases the amount of surfactant on the drop surface, as illustrated in figure \ref{inserthere6}a. 
Increasing the Biot number  leads to a decrease in the surface tension, resulting in a higher deformation with $\approx$25\% increase 
from $\text{Bi}=0$ to $\text{Bi}=10$.

\section{Conclusion\label{s:conclusion}}
In the literature many experimental works  \citep{alvarez2010_lang,alvarez2011_jcis,alvarez2012_jcis,sengupta2019_pre}
show that the transport of bulk surfactant is nonlinearly coupled with drop curvature, 
surfactant physicochemical properties, and external flows. 
Analytical investigation on drop hydrodynamics with surfactant sorption kinetics is challenging due to
the complex nonlinear coupling between surfactant diffusion, sorption kinetics, drop deformation and Maragoni stress.
The numerical method in this study provides a useful tool to quantitatively investigate surfactant exchange between the bulk fluid and the drop. 

We numerically examined the effect of surfactant solubility on the deformation and circulation of a drop under a dc electric field. 
In particular we characterize these effects via the dimensionless transfer parameter ($J)$ and Biot number ($\text{Bi}$).
We showed that surfactant solubility combines with the electric properties of the fluids
in non-trivial ways  to produce rich electrohydrodynamics of a viscous drop with $\chi > 0.8$. 

We first focus on the diffusion-controlled regime in \S~\ref{s:results}.
For $(\sigma_r,\varepsilon_r)$ that corresponds to a clean prolate `A' drop under an electric field (\S~\ref{ss:transientA}), 
surfactant solubility affects both the deformation and flow. In most cases explored (\tikzcircle{2.5pt} in figure \ref{phaseplane}b), the presence of insoluble surfactant 
gives rise to a complete flow reversal (from prolate `A' to prolate `B'). 
Increasing surfactants solubility homogenizes the surfactant distribution on the drop and suppresses the Marangoni stress.
In this case we also observe development of
stagnation points and counter rotating eddies,  
with the counterclockwise eddy taking over with increasing  Biot number. 
Results in \S~\ref{ss:transientA} strongly suggest that the critical $\text{Ca}_E$ for an equilibrium drop shape depends on the solubility, and we are now 
investigating how the critical $\text{Ca}_E$ depends on various parameters.

For $(\sigma_r,\varepsilon_r)$ that corresponds to a clean prolate `A' drop under an electric field (\S~\ref{ss:flow}), 
we find that the surfactant solubility does not affect the drop deformation but does affect the flow pattern.
In this case (small $J$, moderate $\text{Bi}$ and $\text{Ca}_E$) we find that the average surface tension does not vary much with the surfactant solubility
because there is very little net change in total amount of surfactant due to adsorption/desorption. However the spatial variation in $\Gamma$ is sufficient to induce different
flow pattern for the range of electric capillary number we used in the simulations. 
We are now investigating if the above observations hold for stronger electric field strength (larger $\text{Ca}_E$).

For $(\sigma_r,\varepsilon_r)$ that corresponds to a clean oblate drop under an electric field (\S~\ref{ss:deformation}), 
we find that surfactant solubility does not affect
the flow pattern at all ($\MyDiamond$ in figure \ref{phaseplane}b): clean and surfactant-covered oblate drops share 
the same clockwise circulation. However, increasing Biot number further accentuates the strong hydrodynamic flow in oblate drops. 
The resulting enhanced deformation is moderately larger than the insoluble surfactant-covered drop 
cases for $\text{Ca}_E\in[0,1.2]$ (figure~\ref{fig8b}a and figure~\ref{inserthere5}b).

In \S\ref{s:fluxtransfer} we further investigate the drop EHD in the sorption-controlled regime with $J=10$. 
We find that if the drop is unstable at a small $J$, its deformation will grow with a faster rate at a higher $J$ in 
\S\ref{subsec:unstable_drop_dynamics_largeJ}. We also find that increasing the surfactant diffusivity (large $J$) 
suppresses the overshoot in drop deformation dynamics in \S\ref{subsec:transient_overshoot}. 
Moreover, increasing the surfactant solubility homogenizes the surfactant distribution even more and the Marangoni stress is almost completely suppressed 
for $\text{Bi}\ge10$. Under these conditions the drop behaves as a clean drop with a much lower average surface tension. 
Figure~\ref{inserthere5}a shows that the critical $\text{Ca}_E$ is reduced by $\text{Bi}$ and
may reach a fixed constant for sufficiently large surfactant solubility. We are currently investigating this dependence.

\section*{Acknowledgements}
HN acknowledges support from John J. and Char Kopchick College of Natural Sciences and Mathematics at Indiana University of Pennsylvania. 
WFH acknowledges support from Ministry of Science and Technology of Taiwan under research grant MOST-107-2115-M-005-004- MY2.
MCL acknowledges support in part by Ministry of Science and Technology of Taiwan under research grant MOST-107-2115-M-009-016-MY3, and National Center for Theoretical Sciences. 
YNY acknowledges support from NSF under grant DMS-1614863, also support from Flatiron Institute, part of Simons Foundation.


\bibliography{solublesurfactant,surfactantnew_yny}

\appendix
\section{Numerical Implementation\label{appendA}}
\begin{figure}[t]
\begin{center}
\centerline{\includegraphics[keepaspectratio=true,width=5.in]{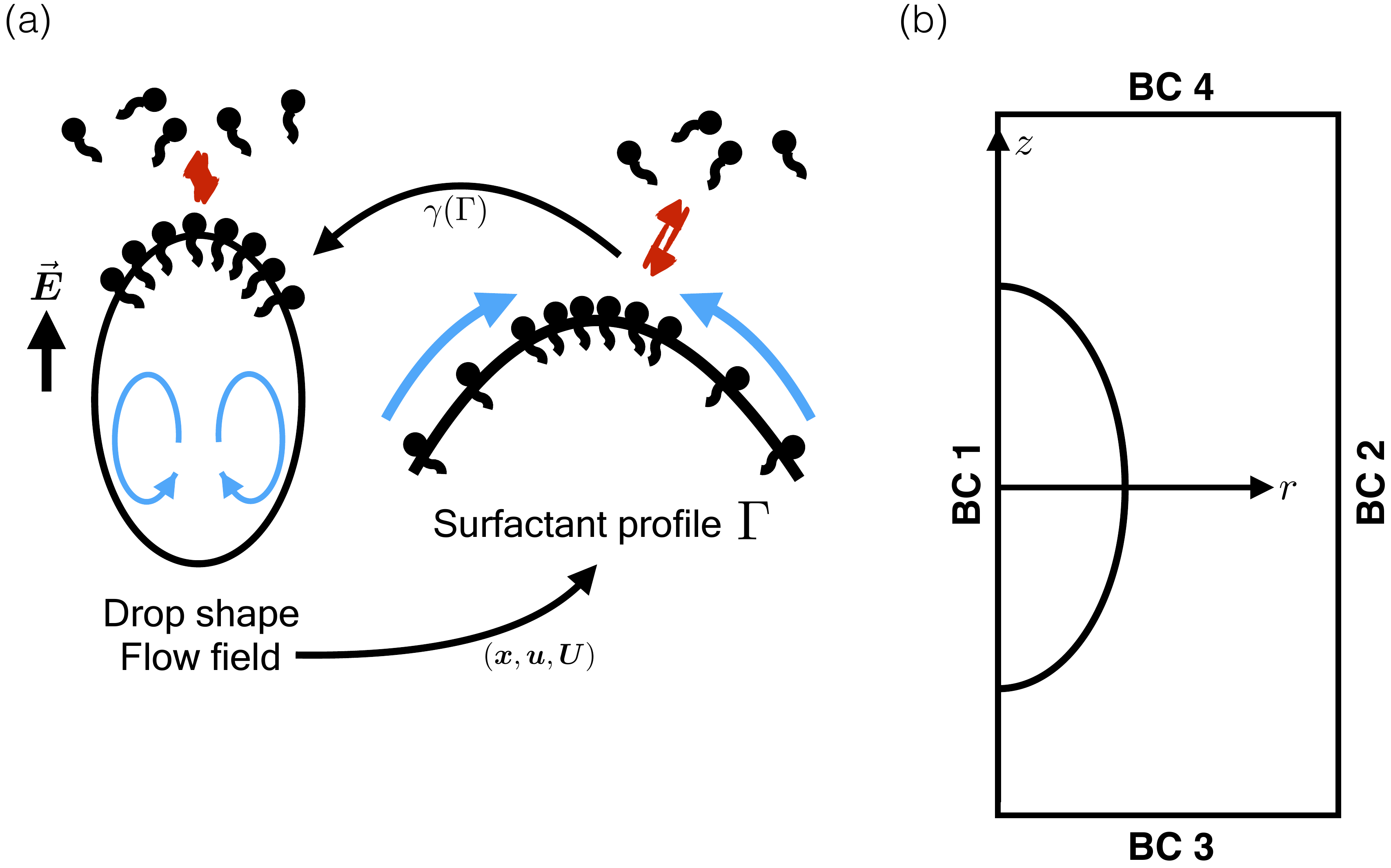}}
\caption{(a) The numerical algorithm for the second-order immersed interface method code: 
At $t_n$ the drop shape $\boldsymbol{x}$, flow field $\boldsymbol{u}$, and interface velocity, $\boldsymbol{U}$ are computed using the electrohydrodynamic solver in \cite{Nganguia2014_CiCP,Nganguia2016_PRE}. The information is then used as input to the surfactant transport solver \cite{hu2018cf}, in order to determine the bulk ($\phi$) and interface surfactant profile ($\Gamma$). 
Given $\Gamma$, we determine the change in surface tension $\gamma$, as well as the updated drop shape, flow field, and interface velocity at time $t_{n+1}$. This process is repeated either until a steady-state is reached, or up to the onset of drop break-up.
Flow circulation and direction are represented by the blue arrows. 
(b) Computational domain on the $(r,z)$-plane. On the walls BC 1, 2, 3, and 4 denote the boundary conditions (see text)
}
\vspace{-7mm}
\label{fig2}
\end{center}
\end{figure}

We solve the governing equations in the axisymmetric cylindrical coordinates $(r,z)$ (figure \ref{fig2}b), considering only the $r\geq0$ half-plane. Once the solution is obtained, it is extended to the left half-plane by symmetry.

Figure \ref{fig2}a illustrates the algorithm. 
The droplet shape and position $\boldsymbol{x}$, flow field $\boldsymbol{u}$ and interface velocity $\boldsymbol{U}$ are computed using the  IIM solver in \cite{Nganguia2014_CiCP,Nganguia2016_PRE}. The boundary conditions in the computational domain $\Omega=[0,L]\times[-L,L]$ in figure \ref{fig2}b are given as follows: for the electric potential, $\phi^+=\mp E_0L/2$ at $z=\pm L$ (the bottom BC3 and top BC4 of the computational domain), while a Neumann boundary condition $\partial\phi/\partial r = 0$ is imposed on the sides ($r=0,\;L$) of the computational domain. For the Stokes equations, the pressure and velocity $\partial p/\partial r=0$, $\partial w/\partial r=0$, $u=0$ at $r=0$ (BC1), while Dirichlet boundary conditions are imposed on the other three sides (BC2-BC4) \cite{Nganguia2014_CiCP}.
For the bulk surfactant concentration $C$, Neumann (BC1) and no flux (zero Neumann) (BC2-BC4) boundary conditions are imposed \cite{hu2018cf}.

For more detailed implementation steps and numerical methods, the reader is referred to \cite{Nganguia2014_CiCP} for the electrohydrodynamic solver. 
The three-dimensional axisymmetric soluble surfactant solver is a straightforward extension of the two-dimensional scheme in \cite{hu2018cf}. The main difference is in the treatment of the correction term for the curvature at the irregular grid nodes. 

\section{Validation\label{appendZ}}
\begin{figure}[ht]
\centering
\includegraphics[keepaspectratio=true,width=5.in]{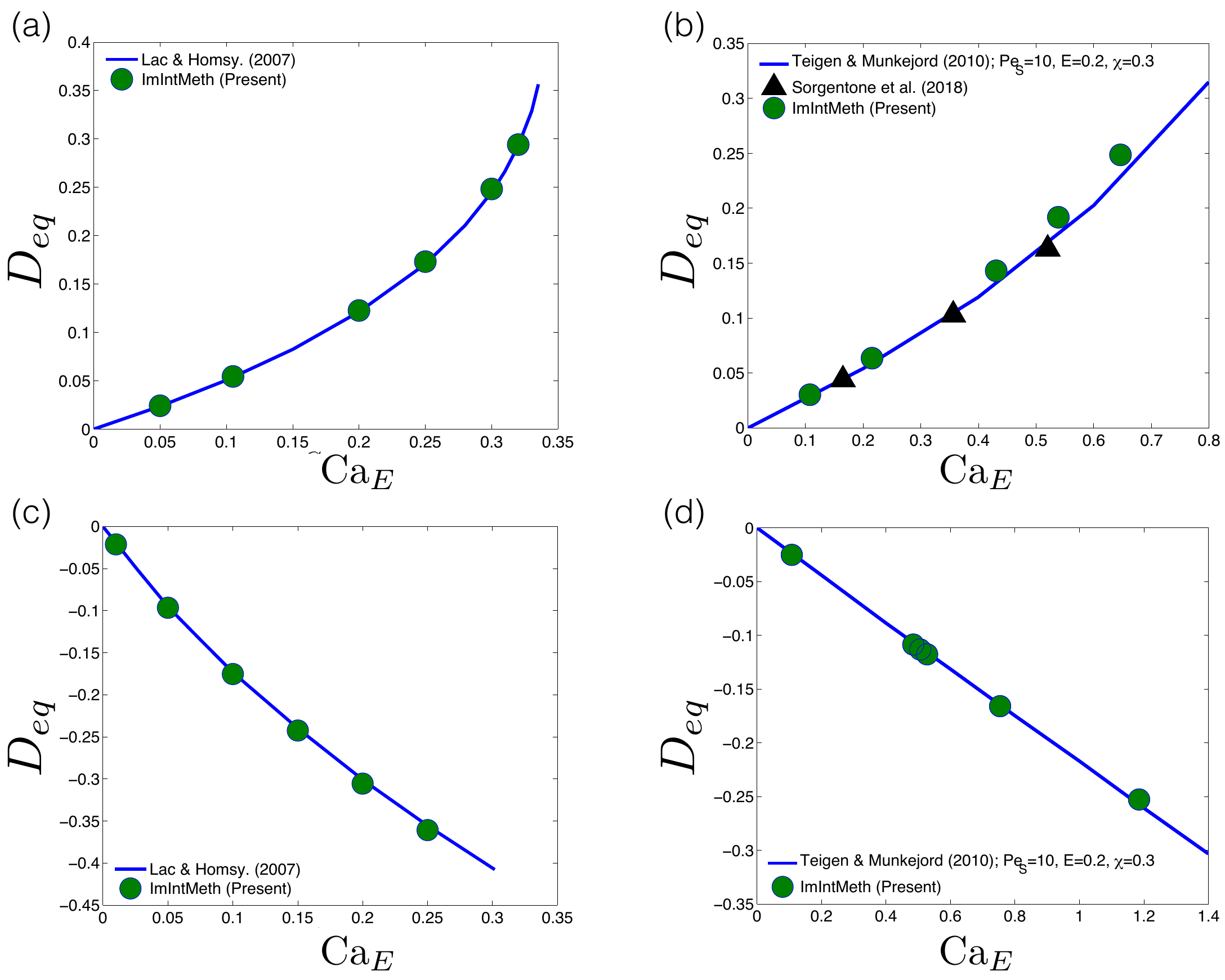}
%
\caption{\small Comparison between published simulation results for the clean drop case (a\&c) in \cite{lh2007jfm} and the surfactant-covered drop cases (b\&d) in \cite{Teigen2010_PhysFluids,sorgentone18_jcp}. The solid lines represent simulations from boundary integral for the clean case, and from level-set for the surfactant-covered drop case. The black triangles represent simulations from boundary integral, while the green circles represent simulations using the proposed immersed interface (IIM) implementation. For the clean drop cases:  
(a) $\sigma_r=0.1,\; \varepsilon_r=0.1$; (c) $\sigma_r=0.5,\; \varepsilon_r=20$.
For the surfactant-covered drop cases we set $E=0.2,\; \chi=0.3,\; \text{Pe}_s=10,\; \text{Bi}=0$:
(b) $\sigma_r=0.3,\; \varepsilon_r=1$; (d) $\sigma_r=1,\; \varepsilon_r=2$. Volume and total surfactant are conserved to within 5\% in all cases. }\label{fig3}
\end{figure}

We validate our numerical codes by comparing against results in the literature where  the equilibrium deformation number (Eq. \ref{eq:Deq})
is reported
as a function of the electric capillary number $\text{Ca}_E$,
for both a clean drop and and a drop laden with insoluble surfactant. $L$ and $B$ are the drop size along the major and minor axes, respectively.
At moderate $\text{Ca}_E$, 
the equilibrium drop shape under a DC electric field could be either
prolate or oblate. For an oblate drop, the circulation is always from the pole to the equator, while the flow inside a prolate drop can be 
ieither from the equator to the pole (prolate `A') or from the pole to the equator (prolate `B').
In our simulations the computational domain size is $[0,5]\times[-5,5]$. The step size $h=5/N$ where $N=256$, and the time step $\Delta t = h/10$.

Figure \ref{fig3} shows comparisons for a clean drop (a\&c) and for a surfactant-covered drop (b\&d).
We test our implementation against the boundary integral (BI) results from figures 5, and 19 in \cite{lh2007jfm}. Figure \ref{fig3}a shows the equilibrium deformation number $D_{eq}$ as a function of the capillary number $\text{Ca}_E$ for a prolate drop with $\sigma_r=0.1,\;\varepsilon_r=0.1$,  
while the oblate drop is shown in figure \ref{fig3}c with $\sigma_r=0.5,\;\varepsilon_r=20$. These comparisons show good agreement with the present immersed interface method (IIM) results.

For the surfactant-covered drop, we consider the work in \cite{Teigen2010_PhysFluids,sorgentone18_jcp} to validate the prolate and the oblate shapes. For these simulations, the electric parameters are set to $\sigma_r=0.3,\;\varepsilon_r=1$ for the prolate drop (case A in \cite{Teigen2010_PhysFluids}), 
and $\sigma_r=1,\;\varepsilon_r=2$ for the oblate drop (case C in \cite{Teigen2010_PhysFluids}). The elasticity constant $E=0.2$ and the surfactant coverage $\chi=0.3$. Other surfactant-related parameters are as follows: the surface and bulk Peclet numbers $\text{Pe}_S=\text{Pe}=10$, respectively, and the Biot number $\text{Bi}=0$ (the insoluble surfactant limit).
Figures \ref{fig3}b and \ref{fig3}d show excellent agreement between all three numerical methods: boundary integral (BI), immersed interface method (IIM), and regularized level-set method (RLSM).

\section{Mesh refinement study\label{appendB}}
We perform a grid analysis (or mesh refinement) study. We consider a computational domain $\Omega = [0,5]\times[-5,5]$, to compute the $L_\infty$ error and determine the ratio 
\begin{equation}
\text{Rate = }\dfrac{||A_{N} - A_{2N}||_\infty}{||A_{2N} - A_{4N}||_\infty},
\end{equation}
where $N$ is the grid size. The number of Lagrangian markers for the interface $M = N/2$. We run simulations to a final time $T=0.5$ with time step $\Delta t = 10^{-3}$. The electric parameters are $\text{Ca}_E=0.1$, $\varepsilon_r=1$, $\sigma_r=0.3$, corresponding to the prolate `A' drop shape (case A in \cite{Teigen2010_PhysFluids}). The surfactant parameters are $E = 0.2$, $\text{Pe}=10$, $\text{Pe}_s=10$, $\chi=0.3$, and the solubility parameter $\text{Bi}=0.01$.
Tables \ref{tab1}-\ref{tab4} show the results of the analysis.

\begin{table}[h]
\caption{Numerical convergence for the flow field variables $\boldsymbol{u} = (u,v)$ and the pressure $p$.}
\vspace{-5mm}
\begin{center}
\begin{tabular}{ccccccc}
$N$ & $||u_N-u_{2N}||_\infty$ & rate & $||w_N-w_{2N}||_\infty$ & rate & $||p_N-p_{2N}||_\infty$ & rate \\ \hline\hline
32 & $1.769\times10^{-1}$ & $-$ & $1.777\times10^{-1}$ & $-$ & $7.303$ & $-$ \\
64 & $2.724\times10^{-2}$ & $2.7$ & $1.086\times10^{-1}$ & $0.711$ & $5.408\times10^{-2}$ & $7.08$ \\
128 & $2.436\times10^{-2}$ & $0.161$ & $4.21\times10^{-2}$ & $1.37$ & $1.402\times10^{-2}$ & $1.95$ \\
256 & $2.498\times10^{-3}$ & $3.29$ & $9.36\times10^{-3}$ & $2.17$ & $2.025\times10^{-3}$ & $2.79$ \\ \hline
\end{tabular}
\end{center}
\label{tab1}
\end{table}%

\begin{table}[h]
\caption{Numerical convergence for the component of interface markers $\boldsymbol{X} = (X,Y)$, the surface surfactant concentration $\Gamma$, the surface tension $\gamma$ and surface tension gradient ${\rm~d}\gamma$. 
}
\vspace{-5mm}
\begin{center}
\begin{tabular}{ccccc}
$M$ & $||\boldsymbol{X}_M-\boldsymbol{X}_{2M}||_\infty$ & rate & $||\Gamma_M-\Gamma_{2M}||_\infty$ & rate \\ \hline\hline
16 & $4.384\times10^{-2}$ & $-$ & $1.053\times10^{-1}$ & $-$ \\
32 & $2.29\times10^{-3}$ & $4.29$ & $9.954\times10^{-3}$ & $3.4$ \\
64 & $4.992\times10^{-4}$ & $2.17$ & $1.72\times10^{-3}$ & $2.53$ \\
128 & $1.07\times10^{-4}$ & $2.2$ & $5.522\times10^{-4}$ & $1.64$ \\ \hline \hline
$M$ & $||\gamma_M-\gamma_{2M}||_\infty$ & rate & $||{\rm~d}\gamma_M-{\rm~d}\gamma_{2M}||_\infty$ & rate \\ \hline\hline
16 & $9.213\times10^{-3}$ & $-$ & $2.27\times10^{-3}$ & $-$ \\
32 & $8.493\times10^{-4}$ & $3.44$ & $1.223\times10^{-3}$ & $0.892$ \\
64 & $1.477\times10^{-4}$ & $2.52$ & $2.897\times10^{-4}$ & $2.08$ \\
128 & $4.744\times10^{-5}$ & $1.64$ & $4.765\times10^{-5}$ & $2.6$ \\ \hline
\end{tabular}
\end{center}
\label{tab2}
\end{table}%

\begin{table}[h]
\caption{Numerical convergence for the staggered variables: the electric potential ($\phi$) and the bulk surfactant concentration ($C$).}
\vspace{-5mm}
\begin{center}
\begin{tabular}{ccccc}
$N$ & $||\phi_N-\phi_{2N}||_\infty$ & rate & $||C_N-C_{2N}||_\infty$ & rate \\ \hline\hline
32 & $1.556$ & $-$ & $2.016$ & $-$ \\
64 & $4.179\times10^{-1}$ & $1.9$ & $1.691$ & $0.254$ \\
128 & $6.779\times10^{-2}$ & $2.62$ & $1.451$ & $0.221$ \\
256 & $1.035\times10^{-2}$ & $2.71$ & $6.025\times10^{-1}$ & $1.27$ \\ \hline
\end{tabular}
\end{center}
\label{tab4}
\end{table}%

\section{Flow fields at high values of the transfer parameter\label{appendJ}}

\begin{figure}[ht]
\centering
\includegraphics[keepaspectratio=true,width=3.7in]{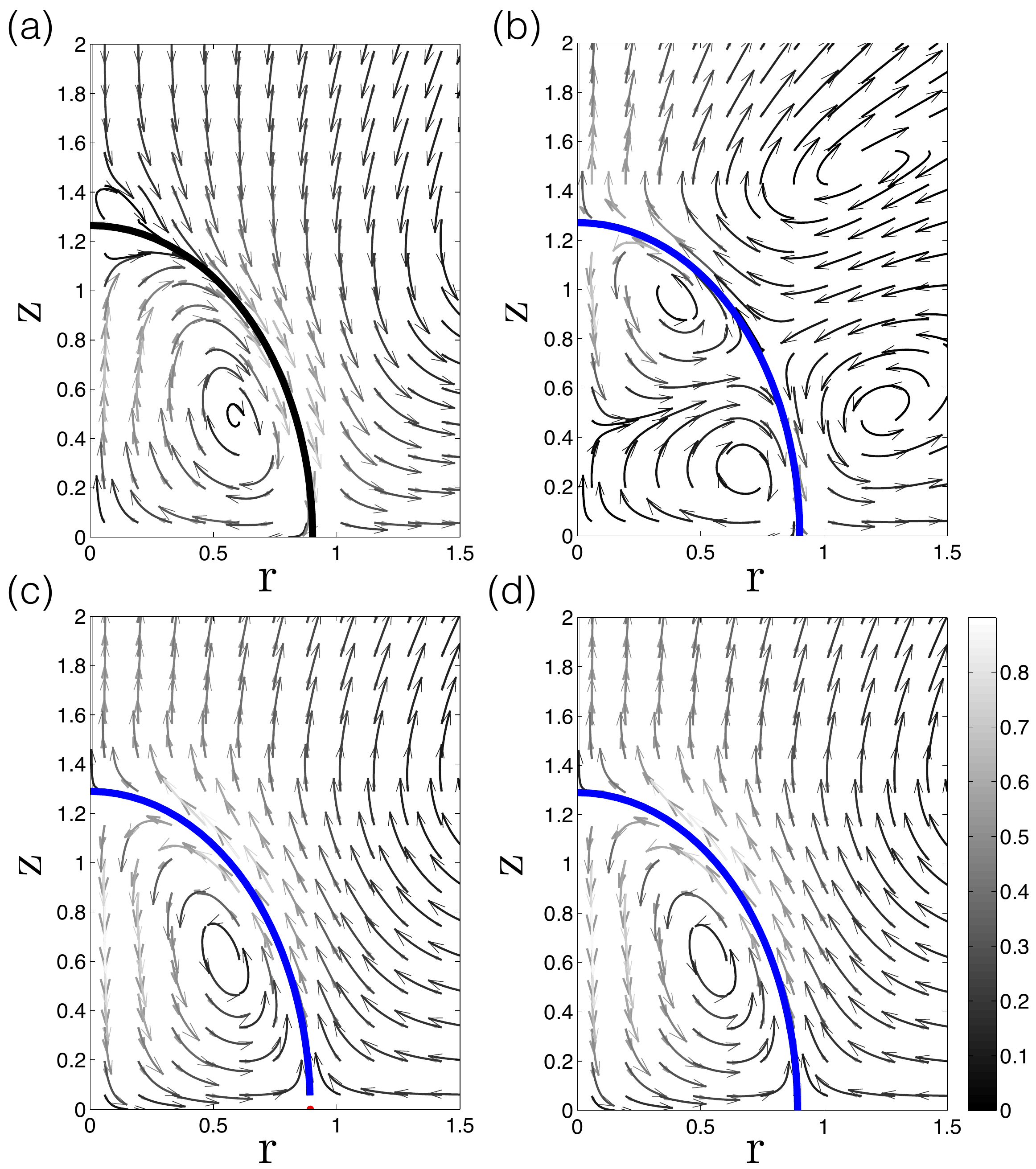}
%
\caption{\small 
Flow field for a prolate drop in figure \ref{inserthere6}a-c with $J=10$. The Biot numbers (a) $\text{Bi}=0$, (b) $\text{Bi}=10^{-2}$, (c) $\text{Bi}=10$, and (d) $\text{Bi}=100$.  The drop surface is color-coded to represent sorption kinetics: blue for adsorption, and red for desorption.
 }\label{insertappend1}
\end{figure}

\begin{figure}[ht]
\centering
\includegraphics[keepaspectratio=true,width=4.7in]{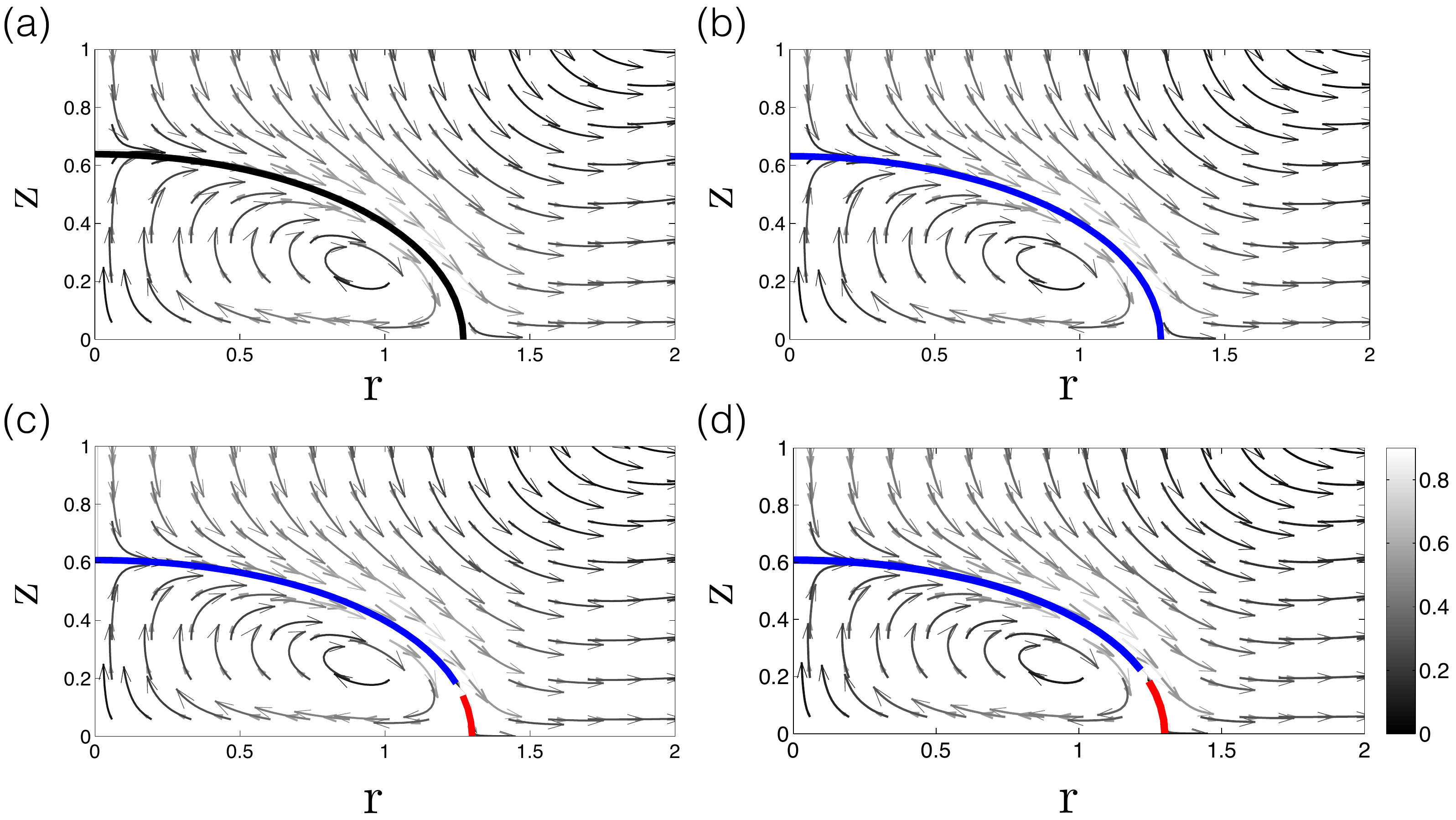}
%
\caption{\small 
Flow field for the oblate drop in figure \ref{inserthere6}d-fi with $J=10$. The Biot numbers (a) $\text{Bi}=0$, (b) $\text{Bi}=10^{-2}$, (c) $\text{Bi}=10$, and (d) $\text{Bi}=100$. The drop surface is color-coded to represent sorption kinetics: blue for adsorption, and red for desorption.
 }\label{insertappend3}
\end{figure}

\end{document}

%% file: definitions.tex
\def\boldsymbol{\bm}